# Leading Edge Vortex Dynamics of Airfoils, Pitching Continuously at High Amplitudes.

Akhil Aravind[1], Pradeep Kumar Seshadri[1,3] and Ashoke De[1,2]

[1]Department of Aerospace Engineering, Indian Institute of Technology Kanpur, Kanpur-208016, India.
[2]Department of Sustainable Energy Engineering, Indian Institute of Technology Kanpur, Kanpur-208016, India.
[3]Iowa Institute of Hydroscience Research, University of Iowa, Iowa City-52246, United States.

*Corresponding Author: ashoke@iitk.ac.in

## Abstract

Airfoils pitching in the stalled regime have been of keen interest in recent years due to their desirable aerodynamic force characteristics. In this study, we are numerically investigating the unsteady flow past a NACA 0012 airfoil under sinusoidally pitching motion, using a finite volume based sharp interface immersed boundary solver. The flow is investigated in the low Reynolds number regime (Re=3000) for reduced frequencies of 0.1 and 0.5, at three different pivot locations (c/3, c/2 and 2c/3 from the leading edge). The airfoil is subjected to sinusoidal oscillations with its incidence angle varying from 15º to 45º. Leading edge vortices (LEVs) that are formed during the pitching motion dictate the transient aerodynamic characteristics. The flow field data is used to identify individual LEVs in the flow field. The spatio-temporal evolution of LEVs in terms of their strengths are traced throughout the pitching cycle to obtain a quantitative estimate of its evolution. The effect of pivot location and pitching frequency on the vortex evolution and aerodynamic forces is investigated. Change in pitching frequency is found to have a drastic effect on the vortex dynamics. LEV formation is delayed to higher fractional times and LEVs are found to interact and merge into each other at higher frequencies. The maximum lift and drag are also found to increase with pitching frequency. Pivot location is found to have a higher influence on the magnitude and phase of aerodynamic coefficients at higher frequencies. However, at both frequencies, moving the pivot axis aftward is found to delay LEV evolution and cause a phase lag on the aerodynamic forces.

## I. Introduction

Biological flyers operate in the low Reynolds number regime where the flow is more likely to separate due to adverse pressure gradient, limiting the maximum steady-state lift. However, these natural flyers enhance their wing's lifting characteristics by flapping it, which generates large-scale unsteady vortical structures called leading edge vortices (LEVs) on them. The high maneuverability of natural flyers is also attributed to these vortical structures. Growing interests in micro-aerial vehicles and aircraft maneuverability have motivated researchers to explore the dynamics of leading-edge vortices and engineer wing motions to take advantage of the transient load these vortical structures impose on the wing.

The motion of the wing, wing cross-section, aspect ratio, freestream conditions and fluid properties influence the dynamics of leading-edge vortices and hence the aerodynamic forces experienced by the wing. Over the years, to isolate the effects of these unsteady vortical structures on the aerodynamic forces on the wing, flow around simple geometries in canonical motions are studied. Most of the studies focus on the overall dynamics and evolution of the leading-edge vortex around a flat plate or a symmetric airfoil under translation[1–4], pitching[5–13] or rotation[14–18]. LEVs in pitching motion were of special interest and were studied for understanding the phenomena of dynamic stall. A brief review of some of the work done in this direction is presented below.

Strickland[5], Walker[6] and Schreck[7] used flow visualization techniques, whereas Visbal[8] used numerical techniques to study the dynamics of NACA 0015 pitching at a constant rate. Shih[9] studied the dynamics of NACA 0012 in pitching motion using both experimental and numerical techniques. Eldredge[10,19] and Jantzen[20] numerically investigated the flow evolution in pitching flat plates, while Granlund[11] and Yu[12] studied the same experimentally. These studies show that the overall dynamics of LEV evolution remains the same irrespective of the cross-section, although; strength, timescale and length scale of LEVs alter with leading edge radius which in turn affects the force histories on the airfoil[21]. Reynolds number (over the typical range investigated: $O(10^2)\ to\ O(10^5)$) is found not to significantly alter the flow dynamics. Studies by Visbal[8], Granlund[11] and Strickland[5] also investigated the effects of pitching frequency and pivot axis on the flow evolution and aerodynamic force coefficients. Increasing the pitching frequency and moving the pivot axis aftward was found to delay LEV evolution, and the maximum lift coefficient was found to increase with pitching frequency. Efforts were also made to develop mathematical expressions to relate the aerodynamic forces to pivot axis location and pitching frequency.

However, most of these studies focus on LEV evolution in a single pitching cycle or half pitch cycle (where the incidence angle goes from minimum to maximum and then remains constant) with simple prescribed wing kinematics. Thus, they do not represent the operation of biological flyers. Most of the studies focus on having their pitch location fixed during the flapping cycle, however studies from Bomphrey[22] show that insects like mosquitoes move their pivot axis between strokes so as to maximize lift in the flapping cycle. The flexible wing structure of these flyers which deform under inertial and aerodynamic loads also tend to enhance their aerodynamic capabilities. Ellignton[23] showed that, for a typical insect flight, the incidence angle varies between 25º and 45º. The flow topology is found to significantly alter with aspect ratio and mean incidence angles. Consequently, smoke-wire visualization of butterfly free flight by Srygley[24] reported generation of LEV pairs (Dual LEVs) on the wing, which are not typical flow structures.

Even without accounting for the complex wing kinematics of birds and insects, Results from Akbari[25] and Ohmi[26,27] show that flow dynamics alter significantly under continuous pitching compared to a single stroke pitching cycle.

Ohmi et al[26,27] used flow visualization techniques to study the flow dynamics around NACA 0012 and elliptical foils pitching in a sinusoidal fashion. He investigated the effects of pitching frequency, pivot axis location, pitching amplitude, mean incidence angle and Reynolds number on the wake patterns. He conducted the studies over the reduced frequencies of 0.1, 0.5 and 1 (reduced frequency is pitching frequency non-dimensionlised with chord length and freestream velocity, $(f^* = fc/2U_\infty)$), while varying the pivot axis location between c/3, c/2 and 2c/3 from the leading edge. The mean incidence angle was varied between 15 and 30 degrees and the angular amplitude was changed between 7 and 15 degrees. The study was conducted for the Reynolds numbers of 1500, 3000 and 10000. He classified the unsteady flows into four wake patterns and identified parametric regimes at which different wake patterns appeared. However, his study was limited to five pitching cycles (at the reduced frequency of 0.5). Streamline patterns from his experiments suggest that, at higher pitching frequencies, a periodicity was not established, and the flow was still transitioning.

Akbari[25] used numerical methods to study the dynamics of NACA 0012 for five complete sinusoidal oscillations. His studies also correspond to Reynolds number varying between 3000 to 10000. He explored the effects of pitching frequency (reduced frequency varying between 0.15, 0.25 and 0.5), pivot axis location (c/4 and c/2) and mean incidence angle (15 and 30 degrees) on the force and moment characteristics of the airfoil. He found that increasing the reduced frequency and moving the pitching axis aftward would delay LEV formation and that the maximum normal force coefficient increases with reduced frequency and mean incidence. Although these results qualitatively correspond to that observed in a half-pitch cycle, the quantitative estimate is expected to be quite different because the streamline contours are found to be significantly different and were found to alter in each cycle. LEVs from different pitching cycles are found to accumulate on the wing at higher reduced frequencies. This can alter the aerodynamic force characteristics in each pitching cycle.

Studies from Zhenyao[28] and Tian[29] show that the phase shift observed in the aerodynamic forces at different pivot axis location can be attributed to the change in the effective incidence angle in a half pitch cycle. Mulleners[30] tried to capture the evolution of dominant vortical structures around sinusoidally oscillating OA209 airfoil at the Reynolds number of 920,000. However, his study corresponds to pitch rate (Pitch rate, $K = 2\pi f^*$) varying between 0.05 and 0.1, where LEV merging and other complex interactions are not encountered. Detailed studies have not been conducted on the quantitative estimate of LEV evolution in continuous pitching motion at high pitching frequencies.

It can be observed from the brief literature survey presented above that understanding the evolution of leading edge vortices is key in understanding as well as modeling the performance of insect and bird flight. But studies so far had focused on parametric space i.e. low frequency, low amplitude and single pitching cycle which are more idealistic than realistic. Hence through our work we attempt to answer two questions. How does the leading edge vortices evolve specifically in the context of an airfoil that performs continuous pitching motion at high amplitudes thus mimicking the real biological flyers? What is the effect of pivot axis and pitching frequency on the leading edge vortex dynamics? We have tracked dominant vortical structures and attempted to shed light on how these parameters affect the aerodynamic force characteristics of the pitching foil. The parameters of the study are chosen based on Ohmi et al's[26] work. A finite volume based sharp interface immersed boundary solver is used for this study[31–33].

## II. Numerical Methods

### A. Solver Details

The flow around the airfoil is resolved by solving the unsteady, 3D Favre averaged Navier Stokes equations using a density based Finite Volume Solver. The solver is pre-conditioned for low Mach number incompressible flows and the governing equations are given below.

$$\Gamma \frac{\partial}{\partial \tau} U + \frac{\partial}{\partial t} Q + \frac{\partial}{\partial x}(E - E_v) + \frac{\partial}{\partial y}(F - F_v) + \frac{\partial}{\partial z}(G - G_v) = S \quad (1)$$

where $\Gamma$ is the preconditioning matrix given by Weiss and Smith[34], Q is the conserved variable vector, E, F, G are inviscid fluxes, $E_v, F_v, G_v$ are viscous fluxes, S is the source term and U is the vector of primitive variables for which the equation set is solved for.

$$U = [p, u, v, w, T]^T \quad (2)$$

Time marching is done using the dual time step approach wherein the physical time steps ($\Delta t$) are discretized using second order backward three-point differencing scheme; and the pseudo time steps ($\Delta \tau$) are discretized using the Explicit Euler differencing scheme. Inviscid fluxes are evaluated using the Advective Upstream Splitting Method (AUSM +) and the viscous fluxes are evaluated using the central differencing scheme. Further details about the solver can be found in the literature[32,33,35].

#### a. Sharp Interface Immersed Boundary Method

A sharp interface immersed boundary framework is employed wherein the flow equations are discretized and solved on a cartesian grid and the airfoil (solid body) is represented by a Lagrangian surface mesh using unstructured triangular elements. To impose the effect of the solid body on the flow field, the flow field equations for the nodes around the surface mesh is reconstructed using an interpolation stencil. The interpolation stencil imposes the necessary no-slip and Neumann pressure conditions at the solid body surface. The nodes along which the interpolation stencil is applied, are called IB (Immersed Boundary) nodes. These nodes separate the fluid nodes from those enclosed by the solid body. The novel framework which is reported by Seshadri and De[31,33] employs the interpolation stencil along the local normal direction which helps in resolving the flow around sharp geometries. This is crucial because vorticity evolution in sharp and rounded trailing edges are found to be significantly different if the sharp edges are not represented accurately. Further details about the immersed boundary framework can be found in the literature[31,35].

#### b. Estimation of Aerodynamic forces

The force on the body is estimated by integrating the surface forces on the body. Since the immersed boundary surface dose not coincide with the cartesian grid points, estimating pressure ($p$) and shear stress ($\tau$) on the solid body surface requires interpolating the values from the neighboring grid points. Force along the $i^{th}$ co-ordinate axis is estimated as,

$$F_i = \oint_\Gamma [-p\delta_{ij} + \tau_{ij}] n_j \, d\Gamma \quad (3)$$

Lift and drag are estimated normal and along the freestream directions, and non-dimensionlised with freestream dynamic pressure and chord length to obtain the lift and drag coefficients, respectively.

$$C_L = \frac{L}{\frac{1}{2}\rho U_\infty^2 c}; \qquad C_D = \frac{D}{\frac{1}{2}\rho U_\infty^2 c} \quad (4)$$

### B. Problem Set-up

The dynamics is studies at a Reynolds number of 3000. NACA 0012 with a chord length of 25mm is set into sinusoidal pitching wherein the incidence angle ($\alpha$) is described by,

$$\alpha(t) = 30^o - 15^o \cos(2\pi f t) \quad (5)$$

where 'f' is the frequency of the pitching motion. The mean incidence angle is 30° and the angular amplitude is 15°. The present study is conducted at the reduced frequencies ($f^* = fc/2U_\infty$) of 0.1 and 0.5, while varying the pivot axis location between c/3, c/2 and 2c/3 distances from the leading edge. The simulations are run for 10 full pitching cycles. The angular orientation of the airfoil along the pitching cycle is shown in **Fig. 1.** For the rest of the discussion, the pitching frequency refers to reduced frequency as defined above, given by Ohmi et al[27].

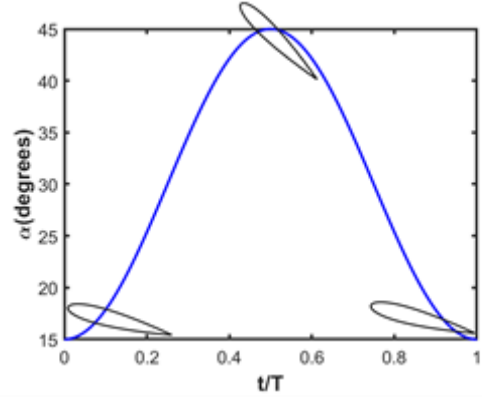

**Figure 1**: Incidence of NACA 0012 along the pitching cycle. The angular orientation of the airfoil is shown at three different instances during the pitching cycle.

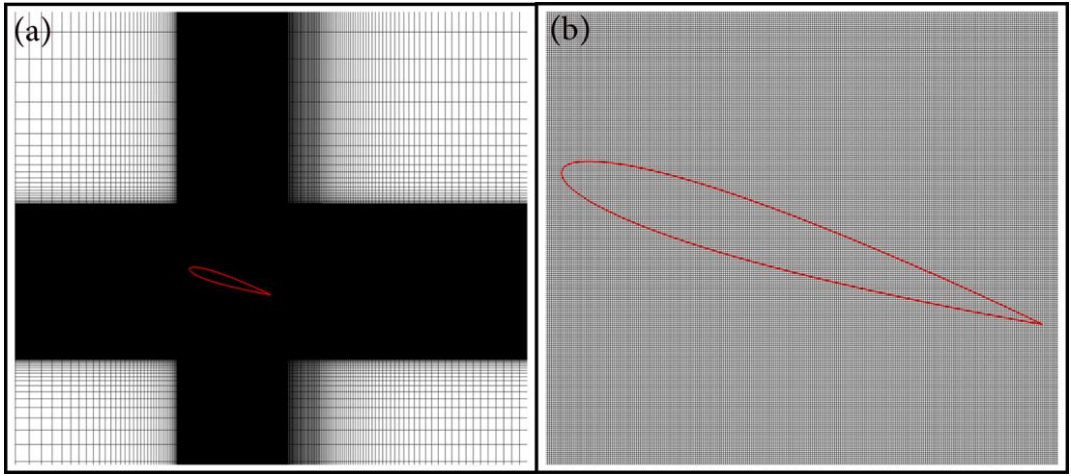

**Figure 2**: (a) Cartesian grid used by the immersed boundary solver. (b) Enlarged view of the cartesian grid around the airfoil

**Fig. 2a** shows the mesh used in the study, and **Fig. 2b** shows an enlarged view of the mesh structure around the airfoil. The mesh structure is 3D, but periodic boundary conditions are imposed along the span-wise direction to make the flow field two dimensional. The grid spacing used around the airfoil is c/200. Validation studies of the case against the experimental data of Ohmi et al[27] can be found in literature[33,35].

## C. Vortex Tracking Method

Dominant vortical structures in the flow field (Leading Edge vortices and Trailing edge vortices) are tracked along the pitching cycle. The idea is to understand the evolution of these vortices in the flow field using a quantitative estimate. With a threshold vorticity of 10% of the maximum vorticity, different vortical structures in the flow field were isolated. The strength of each vortex is calculated by integrating the velocity vector along its boundary to obtain the circulation ($\Gamma$) around it. Circulation is further non-dimensionlised with freestream velocity and chord length ($-\Gamma/(cU_\infty)$).

$$\Gamma = \oint \vec{u}.\vec{dl} \qquad (6)$$

The locus of the vortices was tracked along the pitching cycle by calculating its vorticity weighted centroid given by,

$$x_{cen} = \frac{\oiint_A (\omega x) dA}{\Gamma} \qquad y_{cen} = \frac{\oiint_A (\omega y) dA}{\Gamma} \qquad (7)$$

**Fig. 3** illustrates the dominant vortical structures captured in the flow field. The code tracks the dominant vortices by separating out regions with vorticity greater than 10% of the maximum vorticity. Studies from Shih et al[9] show that varying the vorticity threshold from 10% to 30% of the maximum vorticity does not alter the estimated vorticity weighted centroid.

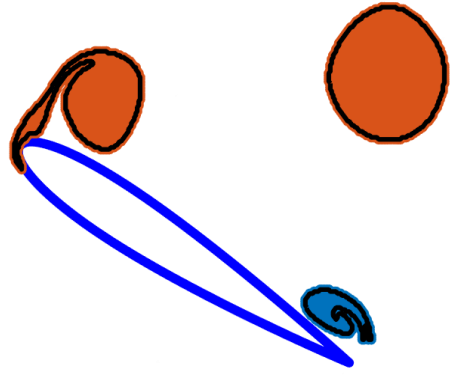

**Figure 3**: Dominant vortical structures captured by vortex tracking algorithm. The boundary of each vortex is traced out in black. The structures filled in orange show clockwise rotating vortices, whereas those filled in blue show anti clockwise rotating vortices.

## III. Results and Discussion

**Section. IIIA** and **IIIB** details the dynamics at the reduced frequencies of 0.1 and 0.5, respectively. A quantitative estimate of LEV evolution is provided in each section along with exploring the effects of changing the pivot axis location on the vorticity evolution and aerodynamic forces. In **Section. IIIC,** we will discuss the dependence of pitching frequency on the aerodynamic forces. More details into the trends observed in the lift and drag characteristics against pitching frequency and pivot axis location can be found in **Section. IIID**.

## A. Dynamics at $f^* = 0.1$

As a baseline case to describe the LEV evolution in the flow field, let us consider the airfoil pitching about its mid chord. **Fig. 4** shows the vorticity evolution around the airfoil during the first pitching cycle. As the pitching cycle begins from its lowest angle of attack, a trailing edge vortex (TEV) is shed **(Fig. 4; t/T=0.1)**. Even the lowest incidence angle is greater than the static stall angle. Boundary layer separation which starts at the trailing edge advances towards the leading edge. The shear layer instability at the leading edge causes the flow to roll up into the leading-edge vortex (LEV) **(Fig. 4; t/T=0.3)**. The influx of mass into the LEV, via the feeding shear layer (FSL) at the leading edge, makes the LEV grow in both size and strength **(Fig. 4; t/T=0.5)**. As the LEV grows, it induces a region of reversed vorticity under it. This vortex structure is referred to as secondary counter rotating vortex (SCV) and can be seen in **Fig. 4 (t/T=0.5).**

As the pitching cycle continues, SCV grows and it tends to obstruct the influx into the LEV. The feeding shear layer which, initially was engulfed by the LEV, starts to appear as a distinct structure at the leading edge. As the motion continues, SCV grows large enough to cut off the link between the LEV and the feeding shear layer. LEV detaches from the leading edge and starts to advect in the wake **(Fig. 4; t/T=0.7)**. The feeding shear layer rolls up another vortical structure at the leading edge. From this point on, the vorticity evolution can happen in two ways. The new vortex either rolls down towards the trailing edge along the top surface of the airfoil **(Fig. 4; t/T=0.8-1.0)** or advects away in the wake along with the LEV **(Fig. 5; t/T=6.7-6.8)**, leaving behind of portion of it to roll down along the top surface of the airfoil to reach the trailing edge **(Fig. 5; t/T=6.8-7.0).**

In the rest of the paper, the former evolution type will be referred to as **'Type-1'** and the latter will be referred to as **'Type-2'**. The first dominant vortex that is rolled up at the leading edge, in each cycle will be described as the **Main LEV** or simply **LEV**. The vortex that rolls down the top surface of the airfoil will be referred to as **Secondary LEV(SLEV)**, and the vortex that is advected along with the main LEV in the Type-2 evolution, will be referred to as **LEV-b**. This classification of LEV evolution is only based on the number of dominant vortical structures in the flow field. Each pitching cycle is found to evolve either in Type-1 or Type-2 pattern. However, the strengths of the dominant vortices can vary from cycle to cycle within a given evolution type.

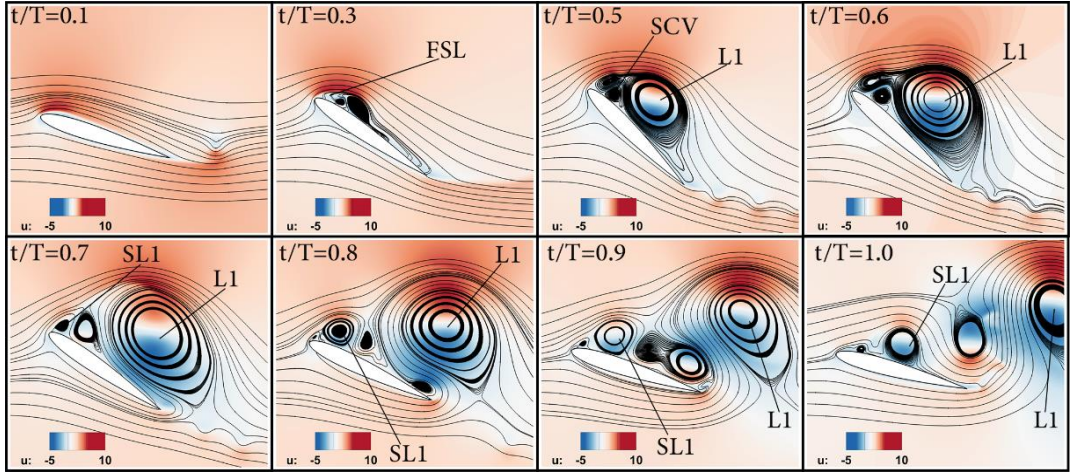

**Figure 4**: Streamline contours in the first cycle of NACA0012 pitching about its mid-chord. LEVs in the figure are indicated by 'L', secondary LEVs are indicated by 'SL', secondary counter rotating vortices are indicated by 'SCV', and the feeding shear layer is abbreviated as 'FSL'. The number following the LEVs describe the pitching cycle in which they first appeared.

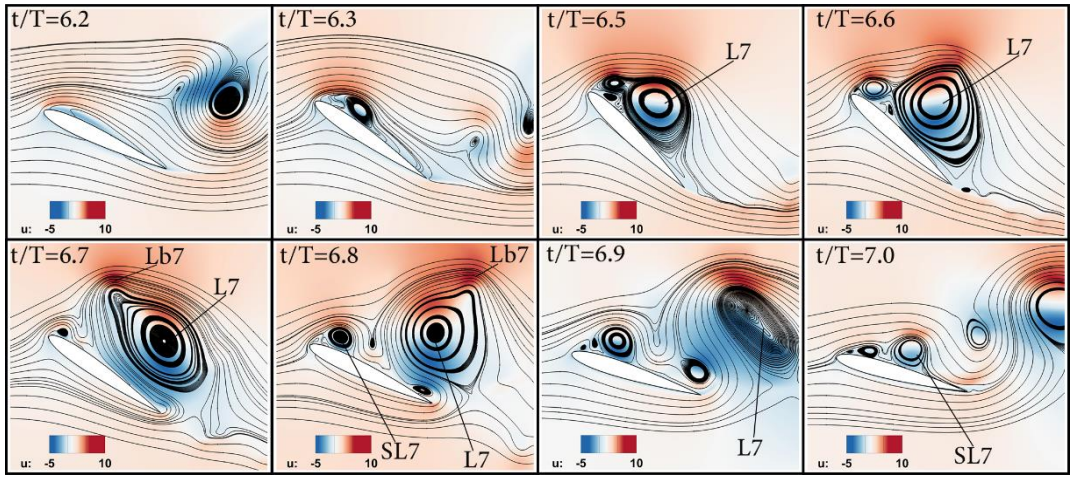

**Figure 5**: Streamline contours in the seventh cycle, for airfoil pitching about its mid-chord. The flow evolution is Type-2.

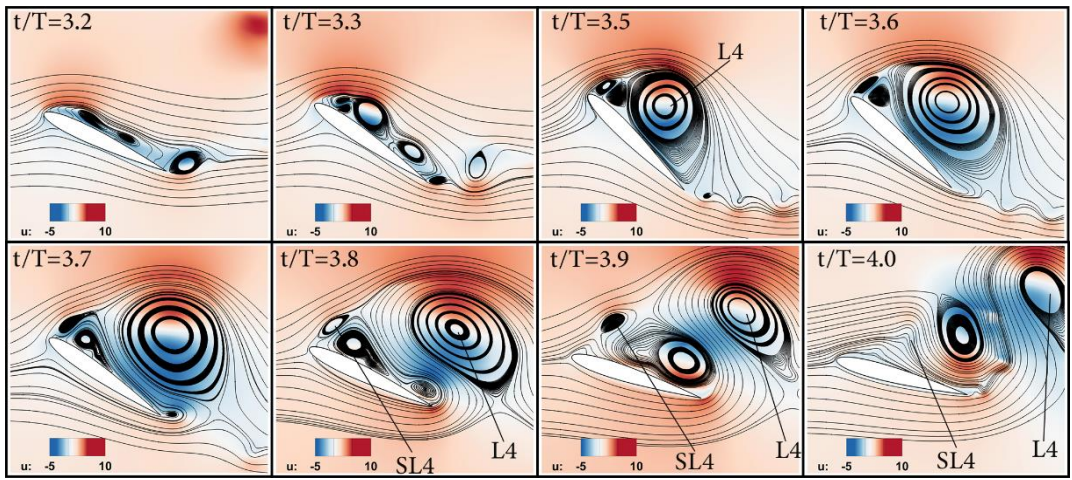

**Figure 6**: Streamline contours in the fourth cycle, for airfoil pitching about its mid-chord. The flow evolution is Type-1, like that observed in the first pitching cycle (**Fig. 4**)).

For the rest of the discussion, LEVs will be identified by 'L', Secondary LEVs by 'SL' and LEV-b by 'Lb'. A number is designated to the vortices along with these alphabets that indicates the pitching cycle in which they originated. For example, L7 refers to LEV generated in the seventh pitching cycle. Likewise, Lb7 refers to LEV-b generated in the seventh cycle.

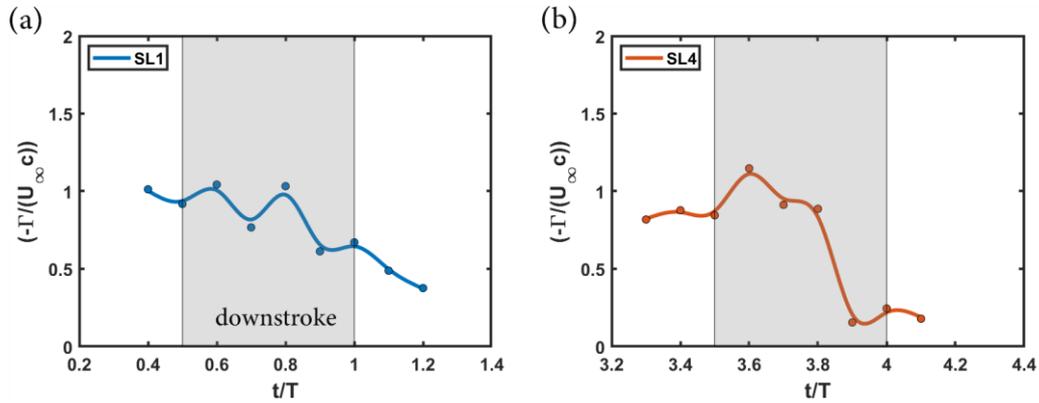

**Figure 7**: Circulation evolution of Secondary LEVs (SLs) along the first (a) and the fourth (b) pitching cycle. The marked points represent the data points obtained from the vortex tracking algorithm, and they are curve fitted to get a smooth plot.

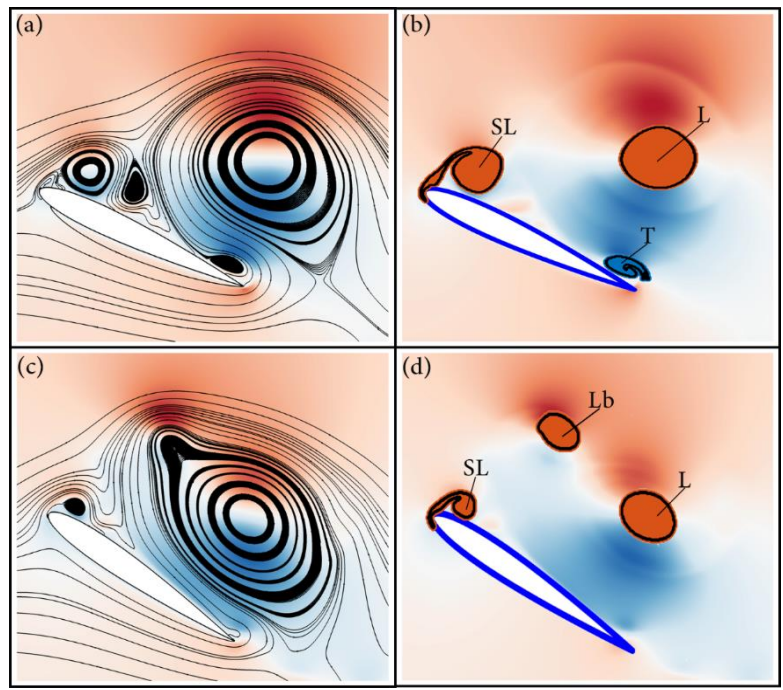

**Figure 8**: (a) and (c) shows the streamline contours corresponding to NACA 0012 pitching at the reduced frequency of 0.1, evolving in Type-1 and Type-2 patterns, respectively. The dominant vortical structures in the flow field corresponding to streamline contours in (a) and (b) are shown in (b) and (d) respectively. In the figure, 'L' represents leading edge vortices (LEVs), 'SL' represents Secondary LEVs and 'Lb' represents LEV-b.

**Fig. 6** shows the streamline plots corresponding to the fourth pitching cycle. The evolution type in the fourth cycle (**Fig. 6; t/T=3.7-4.0**) is same as that observed in the first cycle (**Fig. 4; t/T=0.7-1.0**). The difference in the vortical strengths of the SLs between the two cycles are clearly visible from the streamline contours. **Fig. 7** plots the circulation evolution of the SLs of the first and the fourth pitching cycle. We can see that the circulation (non-dimensionlised) plots are different from each other in the quantitative sense. During the downstroke motion (indicated by the greyed area), the circulation around SL drop at a much faster rate in the fourth pitching cycle compared to the first pitching cycle.

Even though the streamline contours downstream of the airfoil is the same in both evolution types, they differ significantly in terms of dominant vortical structures in the flow field. There are 2 major vortices identified in Type-1 evolution (L and SL), whereas there are 3 such structures identified in Type-2 evolution (L, Lb and SL). **Fig. 8** identifies major vortices in the flow field in Type-1 and Type-2 evolution. **Fig. 8a** and **Fig. 8b** corresponds to Type-1 evolution while **Fig. 8c** and **Fig. 8d** to Type-2.

The vortex tracking algorithm identifies the feeding shear to be a part of the main LEV (L) when it starts forming. However, as the SCV grows, the distinction between them LEV and FSL becomes clear, and the FSL is tracked as a part of the next vortex it rolls up (Secondary LEV (SL) in Type-1 evolution and LEV-b (Lb) in Type-2 evolution).

Next, we will have a look at the quantitative estimate of LEV evolution in Type-1 and Type-2. **Fig. 9** and **Fig. 10** quantifies the evolution of dominant clockwise vortices, by tracking their circulation and vorticity weighted centroid. The figures correspond to the fourth and seventh pitching cycle, which follow Type-1 and Type-2 evolutions, respectively. (Streamline contours corresponding to the fourth and seventh pitching cycles are shown in **Fig. 6** and **Fig. 5,** respectively)

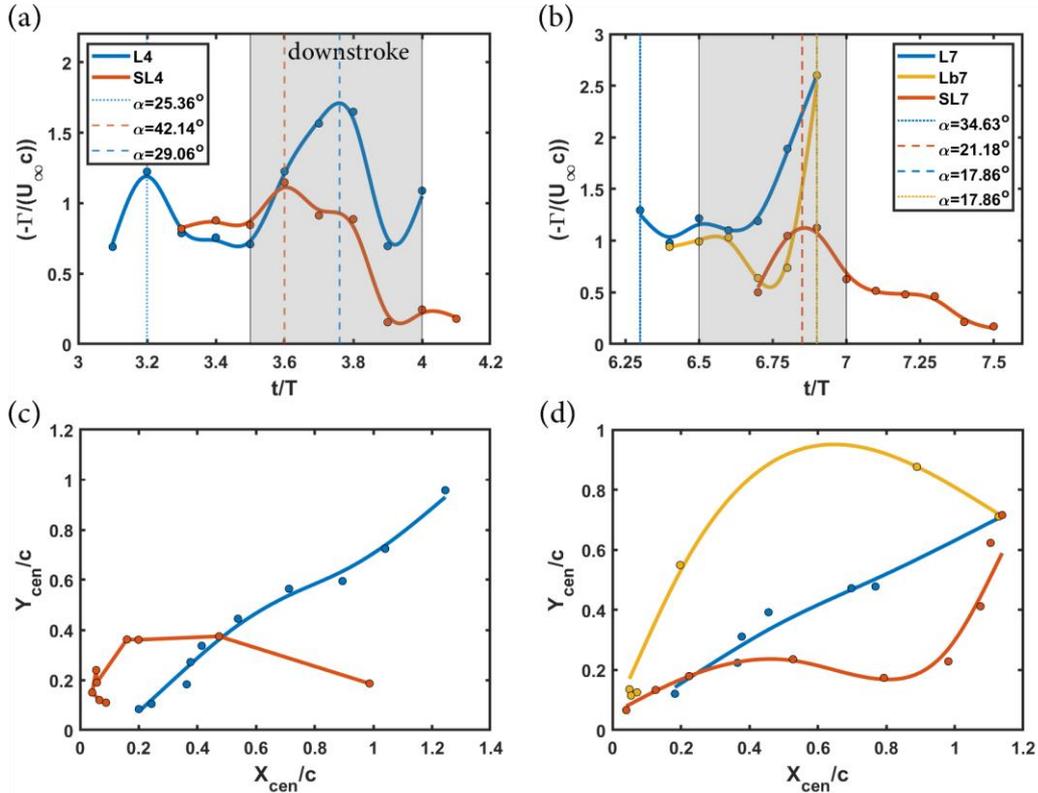

**Figure 9**: (a) and (b) plots non-dimensionlised circulation of the dominant clockwise vortices in the fourth (Type-1) and seventh (Type-2) pitching cycle, respectively. Likewise, (c) and (d) tracks the locus of the vorticity weighted centroid of the LEVs. The shaded region in the plots represent the downstroke motion.

**Fig. 9a** and **Fig. 9b** plots the circulation around dominant vortices in the non-dimensionlised form, for the fourth and seventh pitching cycle, respectively. The main LEVs [L4 in **Fig. 9a and Fig. 9c** (Type-1) and L7 in **Fig. 9b and Fig. 9d** (Type-2)] are traced out in the blue curves. The red curves correspond to the Secondary LEVs [SL4 in **Fig. 9a and Fig. 9c** and SL7 in **Fig. 9b and Fig. 9d**], and the yellow curves (**Fig. 9b** and **Fig. 9d**) trace out the evolution of Lb in Type-2 evolution. The shaded (greyed) and the unshaded region in the plots (**Fig. 9a** and **Fig. 9b**) corresponds to the downstroke and upstroke motion, respectively.

Even though the main LEVs form on the wing at similar incidence angles in the fourth and the seventh pitching cycle, L7 is traced out from a higher fractional time (**Fig. 9b**). This is because the vortex tracking algorithm retains only the regions that has vorticity greater than 10% of the maximum vorticity and L7 tends to exceed this threshold only after the fractional time of 6.3. This corresponds to an incidence angle of 34.63º (upstroke). For both the evolution types (**Fig. 9a** and **Fig. 9b**), the main LEVs show a local maxima in the upstroke motion (t/T=3.2 (25.36º upstroke) in **Fig. 9a** and t/T=6.3 (34.63º upstroke) in **Fig. 9b**), right after its formation. From that point onwards, the distinction between the main LEV and the feeding shear layer becomes evident and the algorithm traces the development of new vortical structures (SL4 in **Fig. 9a** and Lb7 in **Fig. 9b**). Circulation around the main LEVs reach their absolute maxima when the pitching cycle is around 75% (29.06º downstroke) and 90% (17.86º downstroke) complete in terms of fractional time, for Type-1 and Type-2 evolution, respectively. At this point, LEVs are detached from the leading edge but exist on the wing chord.

Lb7 (**Fig. 9b** (Type-2)) which is advected along with the main LEV (L7), attains its peak circulation at the same fractional time as L7. Secondary LEVs attain their peaks around the same when the main LEV (L4 in **Fig. 9a**) is detached from the leading edge in Type-1 evolution, and LEV-b (Lb7 in **Fig. 9b**) is detached from the leading edge in Type-2 evolution. SL4 reaches its maxima around the fractional time of 3.6 (42.14º downstroke), whereas SL7 attains its peak around the fractional time of 6.85 (21.18º downstroke). The peaks values and its corresponding fractional time mentioned above are specific to the fourth and seventh pitching cycle and can vary from cycle to cycle.

The locus of the vorticity weighted centroid of the LEVs in the fourth and seventh pitching cycle is traced out in **Fig. 9c** and **Fig. 9d**, respectively. In the plot, X-axis is aligned along the chord and Y-axis is normal to the chord. The distances are measured with respect to the leading edge of the airfoil and normalized with the airfoil chord length. The main LEVs are found to move away from the airfoil in both chord-normal and chord-wise directions, whereas the Secondary LEVs remain close to the airfoil and move only along the chord. The X-centroid and Y-centroid of the LEVs are separately plotted against fractional time in **Fig. 10.**

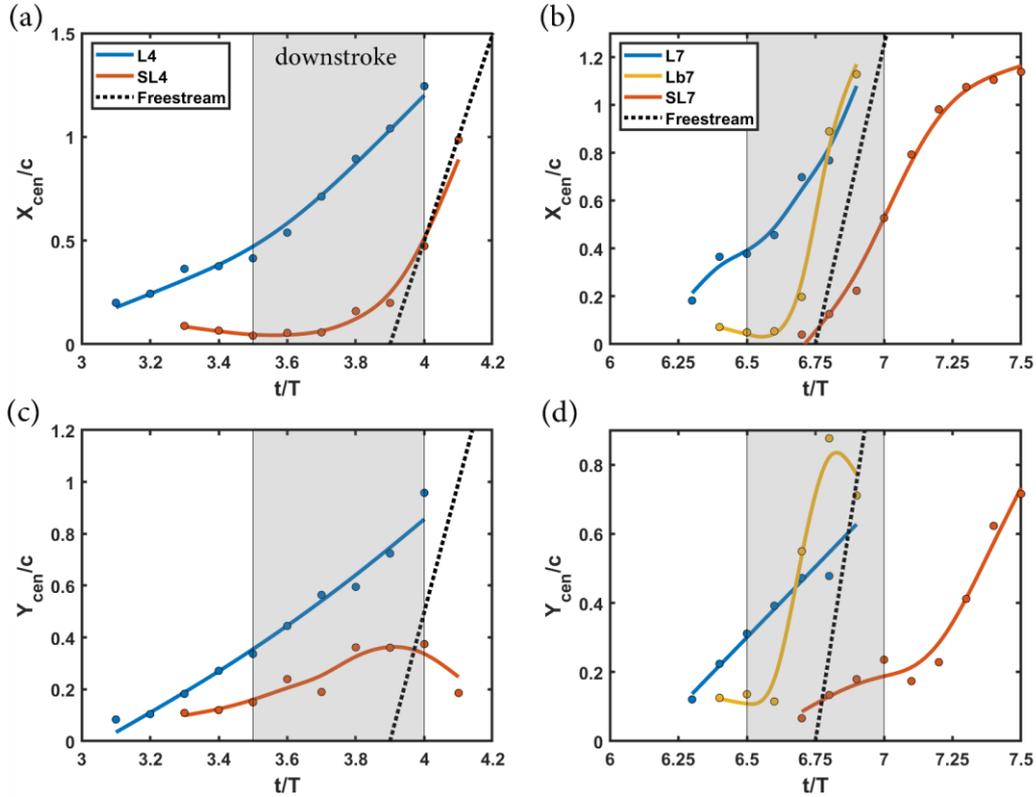

**Figure 10**: (a) and (b) plots the vorticity weighted X-centroid of the LEVs in the fourth (Type-1) and seventh (Type-2) pitching cycle, respectively. Likewise, (c) and (d) tracks the vorticity weighted Y-centroid. The shaded region in the plots correspond to the downstroke motion.

**Fig. 10a** and **Fig. 10b** traces the vorticity weighted X-centroid of the vortices along the pitching cycles. Likewise, **Fig. 10c** and **Fig. 10d** traces the vorticity weighted Y-centroid. For the main LEVs [L4 in **Fig. 10a** (Type-1) and L7 in **Fig. 10b** (Type-2)], X-centroid motion can be approximately broken down into two lines of different slopes indicating that the chord-advection velocity is a piece-wise constant function along the pitching cycle. This transition of velocity happens between fractional times of 3.4 and 3.6 in the fourth pitching cycle and between 6.4 and 6.6 in the seventh pitching cycle. The time intervals correspond to the period when the main

LEVs are separated from the feeding shear layer. When the LEVs are attached to the feeding shear layer, they grow in size and strength. This pushes the vorticity weighted centroid at an almost constant chord-wise velocity. Later, when the LEVs are detached from the leading edge, the vortex starts to advect in the wake and the vorticity weighted centroid attains a higher chord-wise velocity. The vorticity weighted Y-centroid of the main LEVs tend to show a linear variation with fractional time, indicating that the chord-normal velocities of the main LEVs remains constant throughout the pitching cycle. Comparing these velocity estimates with the freestream velocity, we find that both the chord-wise and chord-normal velocities of the main LEVs are lesser than $U_\infty$.

Comparison of the line slopes in Type-2 evolution (**Fig. 10b** and **Fig. 10d**) show that Lb7 attains a much larger advection velocities compared to the main LEV (L7), when detached from the leading edge, in both chord-wise and chord-normal directions (comparable to $U_\infty$). This is also evident from **Fig. 5 (t/T = 6.7-6.8),** where we find Lb7 moving faster than L7.

Although the motion of the main LEVs is comparable in both the cycles, Secondary LEVs do not tend to follow similar trend. SLs, in both evolution types, once detached from the feeding shear layer, starts to roll down the top surface of the airfoil to reach the trailing edge, with chord-wise velocities comparable to the free-stream velocity [SL4 in **Fig. 10a; t/T>3.8** and SL7 in **Fig. 10b; t/T>6.6**]. However, they attain chord-normal velocities of opposite sense in the beginning of the next pitching cycle. Upon reaching the trailing edge, SL7 is advected in the chord-normal direction, which is not observed in SL4. These variations are evident from **Fig. 10c** (SL4; t/T>4) and **Fig. 10d** (SL7; t/T>7). The behavior of SL7 can be accounted for from the corresponding streamline plots, shown in **Fig. 11.** SL7 interacts with a TEV as it reaches the trailing edge. This interaction causes it to advect normal to the chord (positive Y-axis).

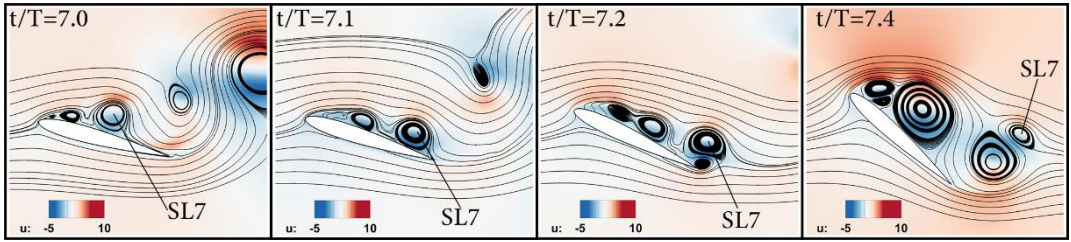

**Figure 11**: Contour plot corresponding to the evolution of SL7 at the beginning of the 8[th] pitching cycle. SL7 interacts with a TEV and attains a chord-normal component of velocity.

Following the vorticity evolution, the aerodynamic forces in Type-1 and Type-2 evolution are expected to be different. The difference is expected in the downstroke motion, wherein the number of dominant vortical structures differ in each evolution type. **Fig. 12a** and **Fig. 12b** plots the lift and drag coefficient, respectively in the fourth (Type-1) and seventh (Type-2) pitching cycles. The plots can be broken down into two curves, the upper envelope corresponding to the upstroke motion and the lower envelope corresponding to the down-stroke motion.

$C_L$ and $C_D$ curves of the two pitching cycles are found to be almost in-phase with each other during the upstroke motion (where maxima are reached), but they differ from each other during the down stroke motion (where minima are reached). Towards the end of the pitching cycle (25° to 15°), both $C_L$ and $C_D$ values are found

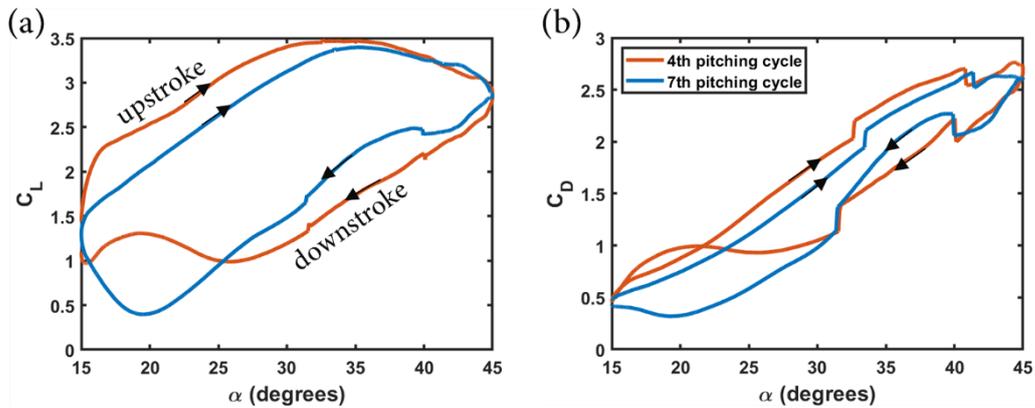

**Figure 12**: Aerodynamic force coefficients in the fourth (Type-1) and seventh (Type-2) pitching cycle is plotted. (a) plots the lift coefficient against incidence angle and (b) plots drag coefficient against incidence

to be higher in the fourth pitching cycle, compared to the corresponding values in the seventh pitching cycle. Since we are comparing the aerodynamic forces between two pitching cycles, for a given case of pitching frequency and pivot axis location, the difference in the lift and drag characteristics must be due to the difference in the vorticity evolution (tracked in **Fig. 9** and **Fig. 10**). The observed differences can be attributed to the additional vortical structure in Type-2 evolution, and all the other small-scale vortices in the flow field that has not been accounted for by the vortex tracking algorithm.

## Effect of Changing Pivot Axis Location (f*=0.1)

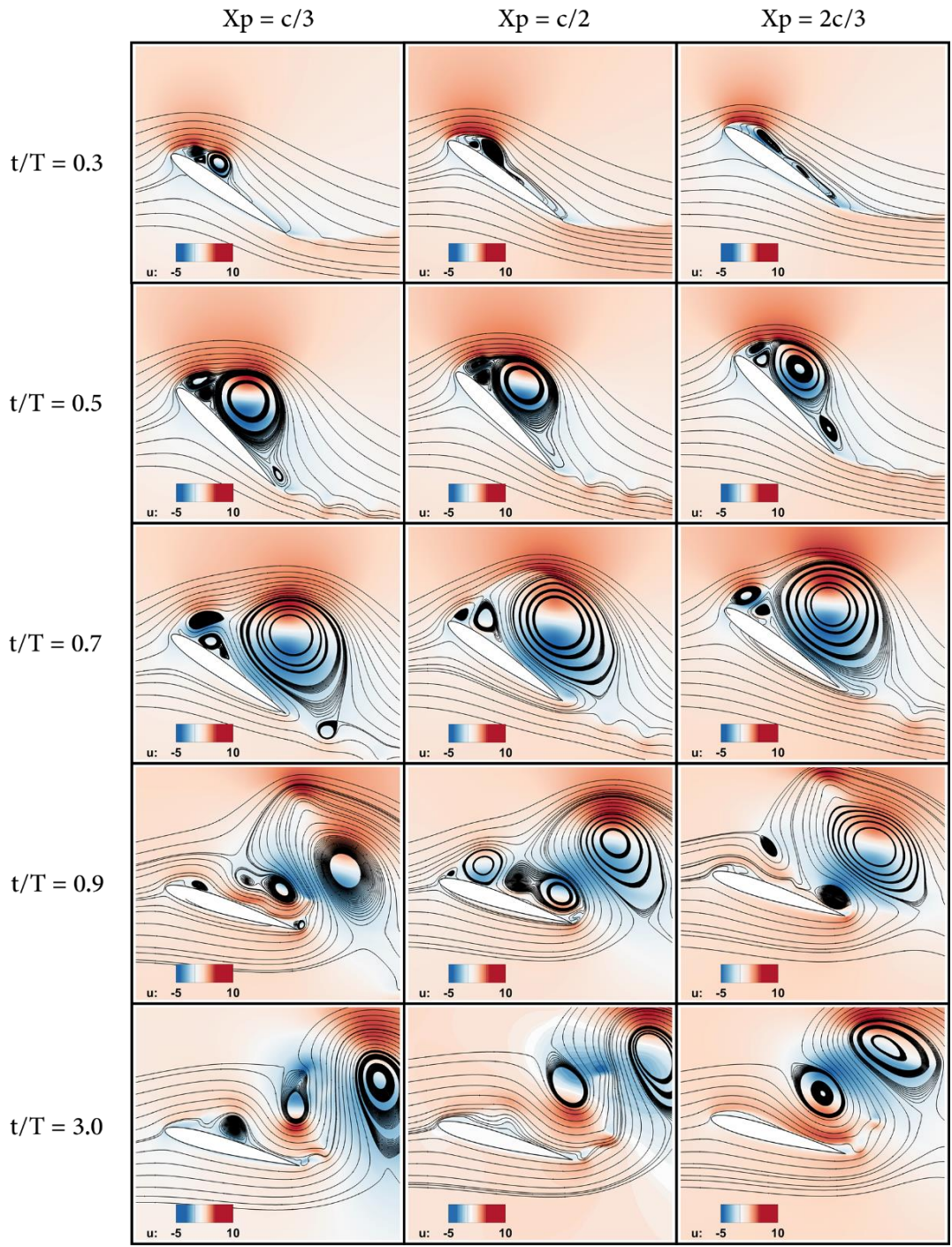

**Figure 13**: Comparison in the vorticity evolution at three different pivot axis locations: c/3, c/2 and 2c/3, ($f^* = 0.1$).

The dynamics of LEV evolution can change with pivot axis location. **Fig. 13** compares the vorticity evolution for three different pivot axis locations at the reduced frequency of 0.1. The pivot axis is varied between c/3, c/2 and 2c/3 distances from the leading edge.

In the first pitching cycle, LEV evolution is Type-1 when the airfoil is pivoted at mid-chord and Type-2 when the airfoil is pivoted at c/3 or 2c/3. Comparing the contours at the fractional time of 0.3, we can see that LEV evolution is delayed as the pivot axis is moved aftward. Eldredge[21] accounted this delay on the basis of chord-normal velocities at the leading edge, given by

$$V_{ch\_n} = U_\infty \left( \sin\alpha - \frac{4\pi f^* x_p}{c} \right) \quad (8)$$

$x_p$ is the pivot axis location with respect to the leading edge. As $x_p$ increases, the chord-normal velocities decrease, which in turn reduces the influx into the LEV, delaying its evolution.

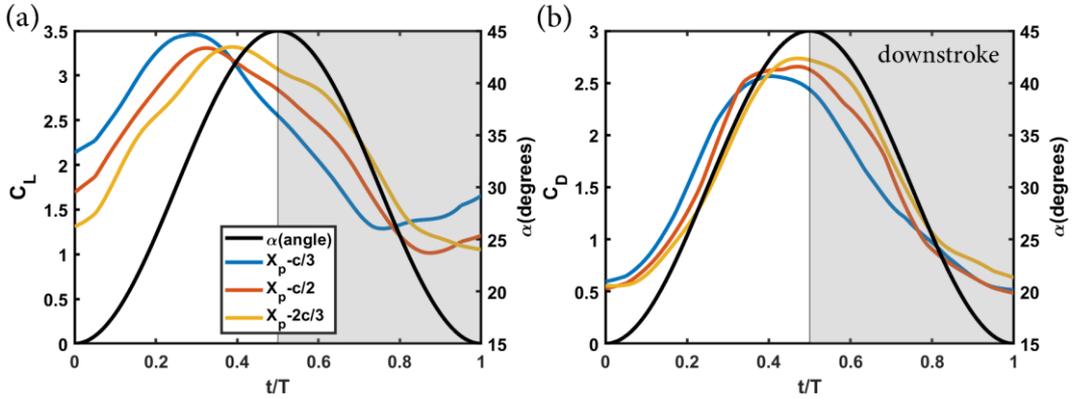

**Figure 14**: Aerodynamic force coefficients on the airfoil compared at three different pivot axis locations ($f^* = 0.1$). (a) plots phase averaged $C_L$ and (b) plots phase averaged $C_D$ against fractional time, respectively.

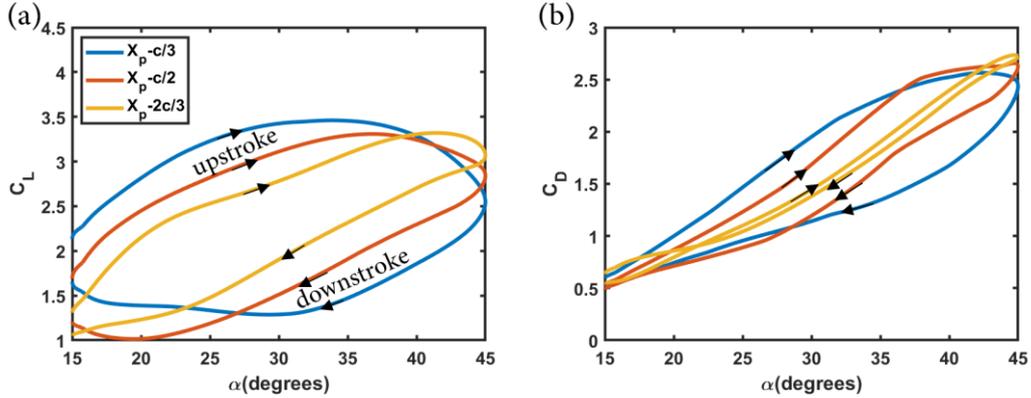

**Figure 15**: Aerodynamic force coefficients on the airfoil compared at three different pivot axis locations ($f^* = 0.1$). (a) plots phase averaged $C_L$ and (b) plots phase averaged $C_D$ against angle of attack, respectively.

However, other than the phase lag observed in the evolution of LEVs as we move the pivot axis aftward, the streamline contours are comparable (even at t/T=3.0). Thus, large differences are not expected in the aerodynamic forces as well. **Fig. 14a** and **Fig. 14b** plots the phase averaged (10 cycles) variation of $C_L$ and $C_D$ against fractional time, respectively. From **Fig. 14a**, we can see that the phase angle at which maxima is reached increases as we move the pivot axis aftward. The observation is true for both lift and drag characteristics. The curves, irrespective of their pivot axis location, leads the motion signal (AOA). $C_{Lmax}$ is found to be highest when the pivot axis location is at c/3 and least when pivoted at mid-chord, with a difference of 4.6% between the two cases. $C_{Dmax}$ is found to increase with distance of the pivot axis location from the leading edge. More trends observed in the aerodynamic forces are discussed in detail in **Section. IIID.**

**Fig. 15** plots the aerodynamic force coefficients against angle of attack, with **Fig. 15a** and **Fig. 15b** corresponding to lift and drag, respectively. Like in **Fig. 12**, the upper envelope of the hysteresis curve represents the upstroke motion, and the lower envelope corresponds to the downstroke motion. The width of the hysteresis curve ($C_L$ and $C_D$) is found to reduce as we move the pivot location aftward. Both $C_L$ and $C_D$ values are found to remain positive throughout the pitching cycle.

Thus, pivot axis is found not to influence the amplitude of the aerodynamic forces at the reduced frequency of 0.1. However, the phase lag observed as the pivot axis is moved aftward is significant. This is synonymous with the observed delay in the LEV evolution as we move the pivot axis aftward (**Fig. 13; t/T=0.3**). Zhenyao[28] and Tian[29] traced out the source of the phase lag to the effective incidence angle. A phase lag like that observed in the aerodynamic loads exists in the effective incidence angle as we move the pivot axis aftward. With respect to mid-chord pitching, the effective incidence angle leads by 0.023T (T is the time period of the pitching cycle) when the airfoil is pivoted at c/3 and lags by 0.023T when pivoted at 2c/3. Details on estimating the effective incidence angle can be found in **Appendix. A1.**

## B. Dynamics at $f^* = 0.5$

LEV evolution is significantly different at the reduced frequencies of 0.1 and 0.5. Let us consider, mid-chord pitching as a baseline case to describe the flow evolution at the reduced frequency of 0.5 (**Fig. 16**).

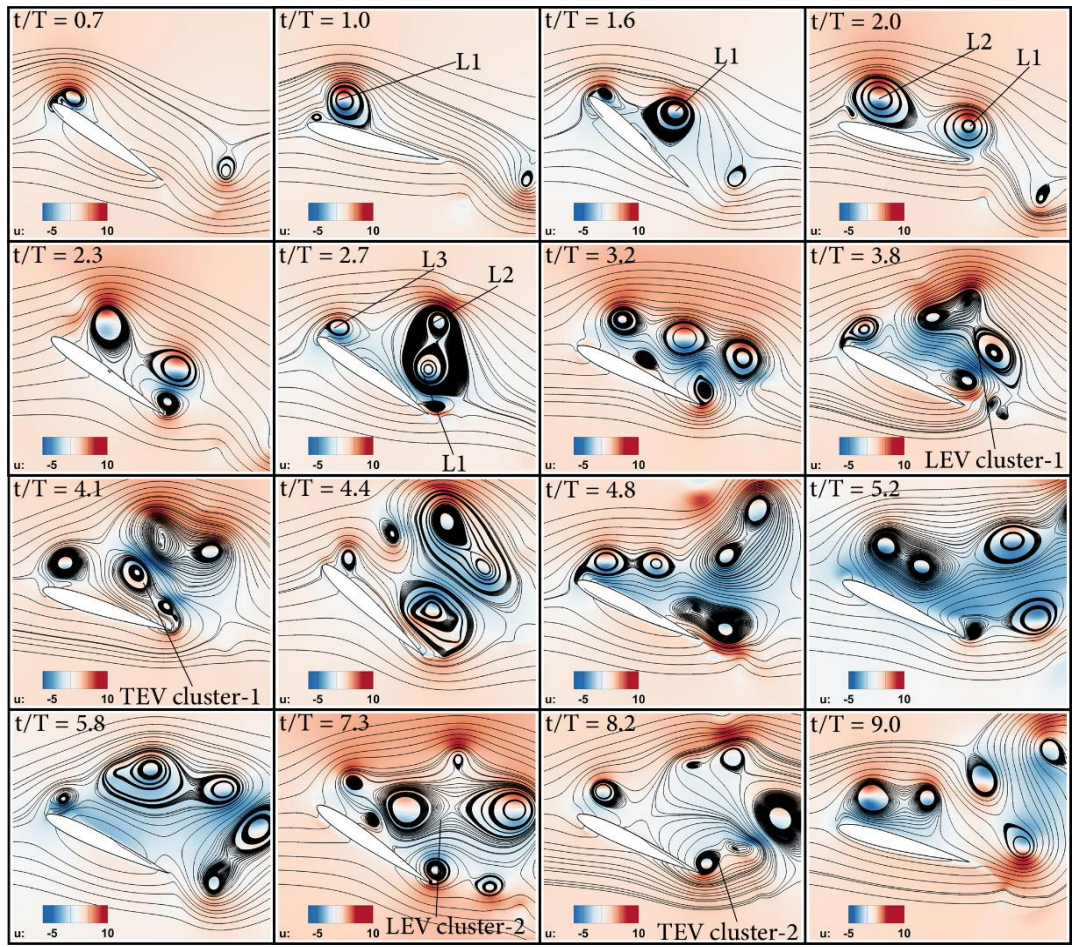

**Figure 16**: Contour plots corresponding to NACA 0012 pitching at the reduced frequency of 0.5, about at its mid-chord.

LEV formation seems to be delayed to larger fractional times at the reduced frequency of 0.5 (in comparison with that observed at $f^* = 0.1$). This again can be attributed to reduction in the chord-normal velocity at the leading edge due to increase in pitching frequency as per **equation 8.**

The pitching motion begins with an upstroke from the lowest angle of attack, during which no large scale vortical structures are observed. Traces of LEV starts to appear during the start of the downstroke motion. By the time the LEV is formed, the TEV generated in the first cycle is shed (**Fig. 16; t/T=0.7**). The LEV formed (L1) grows in size and strength, gets detached from the leading edge and starts to move towards the trailing edge along the top surface of the airfoil (**Fig. 16; t/T=1.0-2.0**).

L1, upon reaching the trailing edge, interacts with the TEV of the next cycle (**Fig. 16; t/T=2.3**). This interaction slows down the advection of L1. During this period, a new LEV is formed on the wing (L2) (**Fig. 16; t/T=2.0**). L2 grows in size and strength, detaches from the leading edge and advects downstream. L2 then interacts with the L1 at the trailing edge (**Fig. 16; t/T=2.7**). This trend continues and LEVs generated in subsequent pitching cycles interact with each other and form an LEV cluster (LEV cluster-1) (**Fig. 16; t/T=3.2-4.1**).

During this period, TEVs formed in multiple cycles also merge with each other (**Fig. 16; t/T=2.7-4.1**) and form a TEV cluster. Finally, this cluster disintegrates (**Fig. 16; t/T=4.4**) and a part of it moves against the freestream direction and breaks the interaction between a new LEV and the LEV cluster. LEV cluster, then begins to break apart and a portion of it gets advected in the wake. The rest of the LEV cluster remains on the wing and interacts with the LEVs of subsequent cycles (**Fig. 16; t/T=4.8-7.3**) forming a new LEV cluster (LEV cluster-2). In the meantime, the TEV cluster is completely detached from the trailing edge and starts to advect in the wake. The cycle repeats. TEVs formed in different cycles merge with each other forming a TEV cluster (TEV cluster-2) (**Fig. 16; t/T=8.2**). Both LEV and TEV cluster grows in size and strength, become unstable and finally breaks apart.

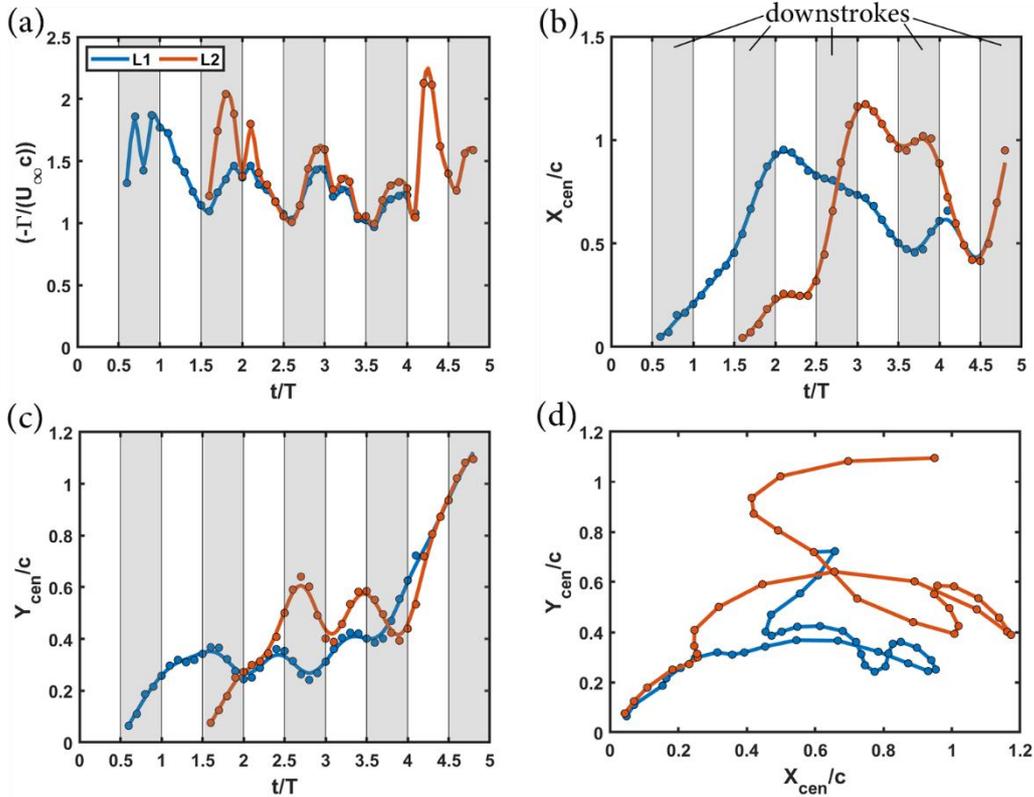

**Figure 17**: Quantitative estimate of LEV evolution ($f^* = 0.5$). (a) plots non-dimensionlised circulation of L1 and L2 along the pitching cycle. (b) and (c) plots vorticity weighted X-centroid and Y-centroid of the LEVs, respectively. (d) tracks the locus of the vorticity weighted centroid of the LEVs along the pitching cycle. The greyed region in the plots correspond to the downstroke motion.

**Fig. 17** plots the quantitative evolution of two typical LEVs, L1 and L2 (see **Fig.16** for the contour plots) in the flow field at the reduced frequency of 0.5. During the initial cycles, (**Fig.17a**) we can see that the strength of L1 and L2 increases during the downstroke motion and decreases during the upstroke motion. This is because, new LEVs are formed during the downstroke motion, and they interact with L1 and L2. However, on an average, these vortices have circulation greater than that observed for typical LEV at the reduced frequency of 0.1.

The variation of their vorticity weighted X-centroid is traced out in **Fig. 17b** and Y-centroid in **Fig. 17c**. From these plots it can be observed that during each upstroke motion, the LEVs are displaced in the chord-normal direction, away from the airfoil, and they return during the downstroke motion (**Fig. 17c**). The locus of the vorticity weighted centroid of the LEVs are traced out in **Fig. 17d**. It can be seen that L1 and L2 go back and forth to remain on the wing till the end of the fifth pitching cycle, when they are finally advected out due to the disintegration of LEV cluster-1.

**Effect of Changing Pitching Axis Location (f*=0.5)**

Flow evolution is found to significantly alter with pivot axis location at the reduced frequency of 0.5. **Fig. 18** shows a comparison of the streamline contours corresponding to three different pivot locations: c/3, c/2 and 2c/3, at the reduced frequency of 0.5. The contours are compared at the end of the second to the sixth pitching cycle. They are found to be significantly different from each other. This can be attributed to timescales and length scales of the vortex cluster, which significantly alter with pivot axis location.

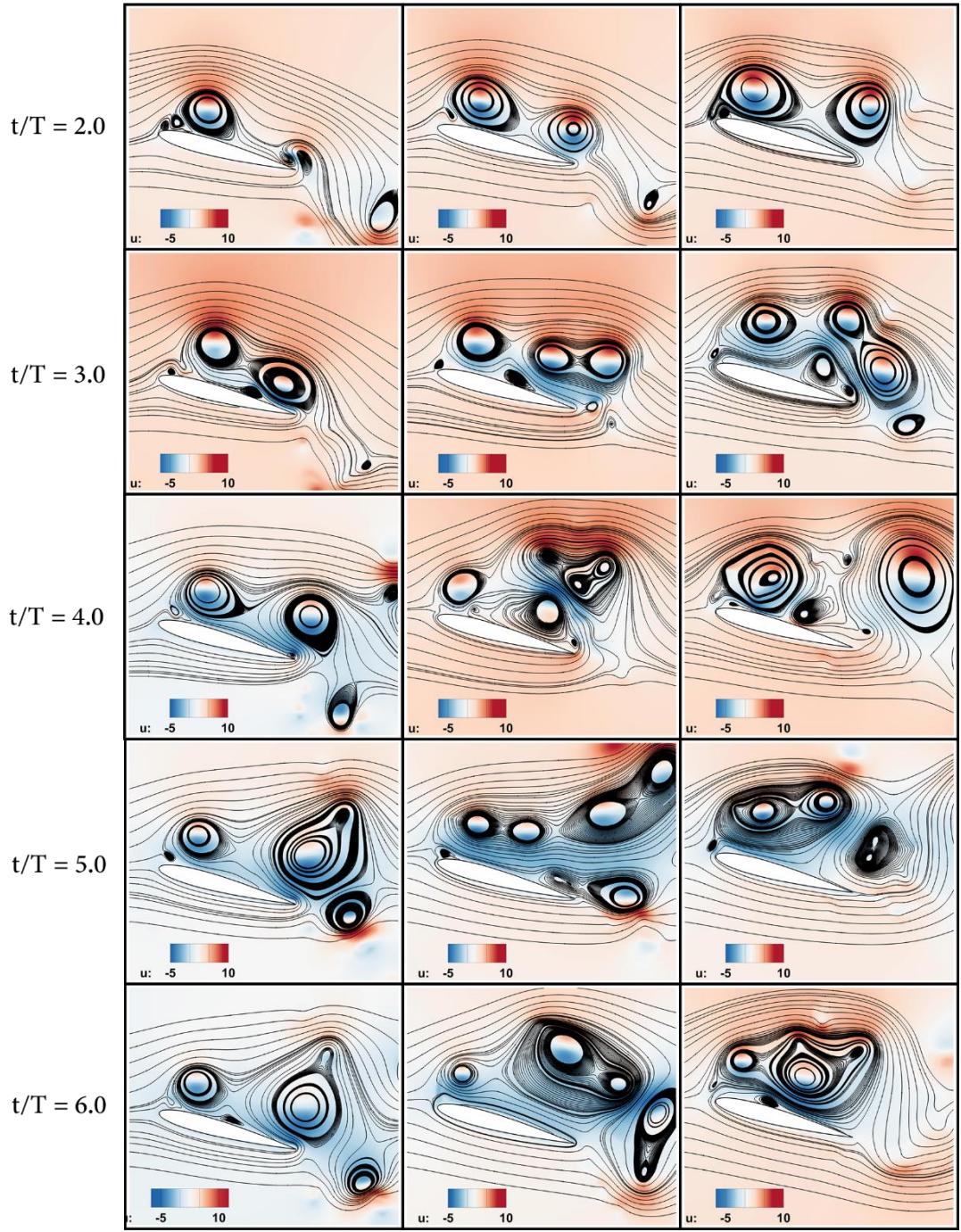

**Figure 18**: Comparison of the streamline contours corresponding to three different pivot axis locations: c/3, c/2 and 2c/3 ($f^* = 0.5$).

**Fig. 19** compares the streamline contours at the fractional time of 0.6 (first cycle). We find that LEV evolution is delayed to higher times as the pivot axis is moved aftward, similar to our observation at the reduced frequency of 0.1 (**Section. IIIA**).

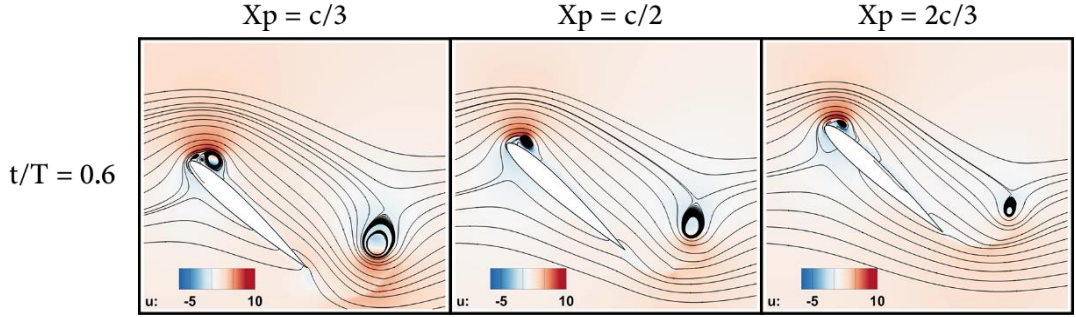

**Figure 19**: Comparison of the streamline contours corresponding to three different pivot axis locations: c/3, c/2 and 2c/3, at the fractional time of 0.6 ($f^* = 0.5$).

Compared to the vorticity evolution described in **Sec. IIIB** for mid-chord pitching, an interesting variant of LEV evolution is observed for one-third chord pitching. The LEV generated in the first cycle merges with the LEV of the second cycle, and the combined structure tends to stay on the airfoil for all the 10 pitching cycles (The flow was simulated for 10 pitching cycles). The contour plots of a few instances corresponding to the evolution pattern is shown in **Fig. 20**. When a new LEV is created, it interacts with the merged vortical structure of L1 and L2 (represented by 'L') that is already present on the wing. The interaction causes the merged structure of L1 and L2 to go back and forth on the wing in such a way that the new LEV is shed in the wake [**Fig. 20; t/T=4.3-4.8 and t/T=5.2-5.7**] and the merged structure is retained on the wing. **Fig. 21** plots the evolution of the merged vortical structure against fractional time.

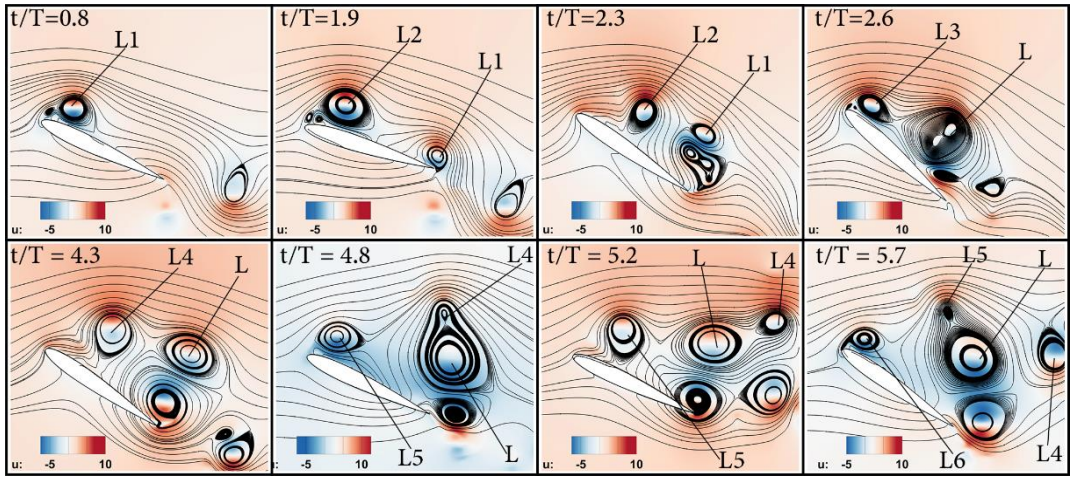

**Figure 20**: Contour plot corresponding to NACA 0012 pitching at the reduced frequency of 0.5, about the one-third chord axis. The merged vortical structure of L1 and L2 is represented by 'L'. Other LEVs are designated by the cycle number in which they first appeared.

Similar to the observation made in **Fig. 17**, the strength of the merged vortical structure 'L' decreases during the upstroke motion and increases in the downstroke motion (**Fig. 21a**). The motion of LEVs (**Fig.21b-c**) are found to follow a periodic pattern repeating in every pitching cycle. The LEVs move towards the leading edge and gets displaced in the chord-normal direction during the upstroke motion. The displacement is reversed in the downstroke motion. **Fig. 21d** traces the locus of the vorticity weighted centroid of the LEVs along the pitching cycle.

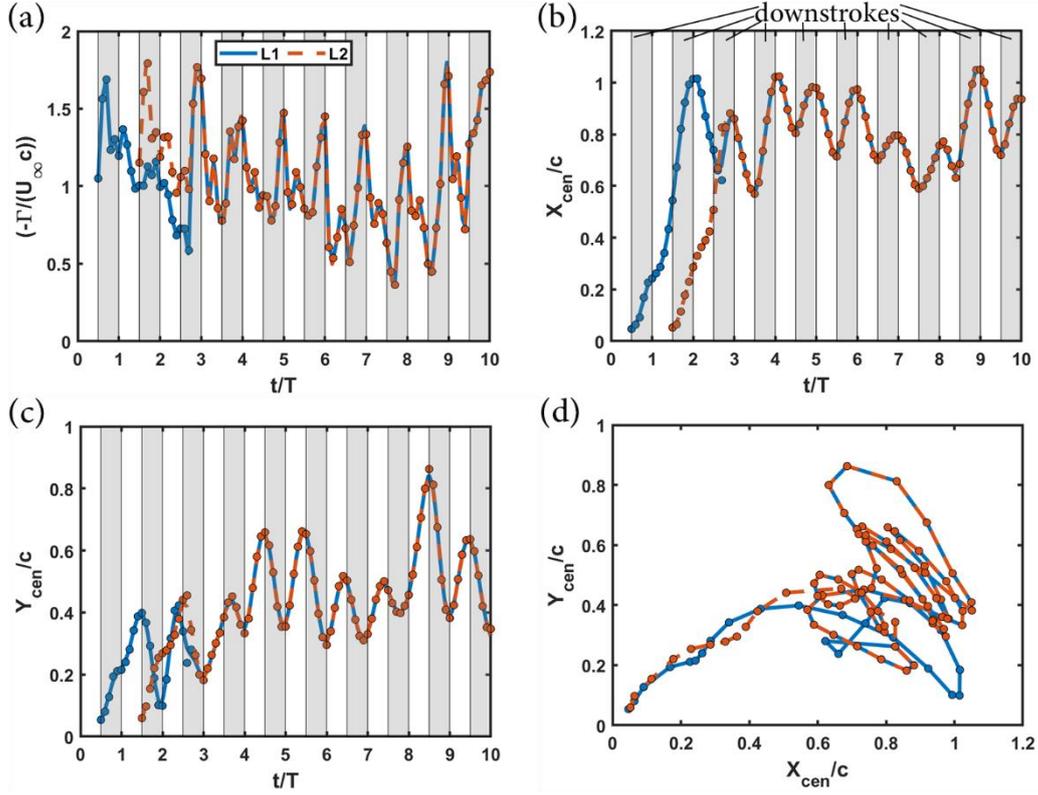

**Figure 21**: Quantitative estimate of L1 and L2 evolution ($f^* = 0.5$), for an airfoil pitching about its one-third chord axis. (a) plots non-dimensionlised circulation of L1 and L2 along the pitching cycle. (b) and (c) plots vorticity weighted X-centroid and Y-centroid of the LEVs, respectively. (d) tracks the locus of the vorticity weighted centroid of the LEVs along the pitching cycle.

The difference in the streamline contours at different pivot axis location suggests a significant difference in the force characteristics. **Fig. 22a** and **Fig. 22b** plots the variation of phase averaged $C_L$ and $C_D$ along fractional time, respectively. The curves corresponding to pivot locations of c/3 and c/2 are found to lead the motion signal, whereas the curves corresponding to pivot location of 2c/3 is found to slightly lag behind the motion signal. The fractional time at which the curves attain their maxima is found to increase as we move the pivot axis aftward. $C_{Lmax}$ is found to be highest when the pivot axis is at c/3 and least when pivoted at mid-chord, with a difference of 62.77% between the two cases. $C_{Dmax}$ is found to be highest for the pivot axis location of 2c/3 and least when pivoted at c/2, with a difference of 30.5% between the two cases.

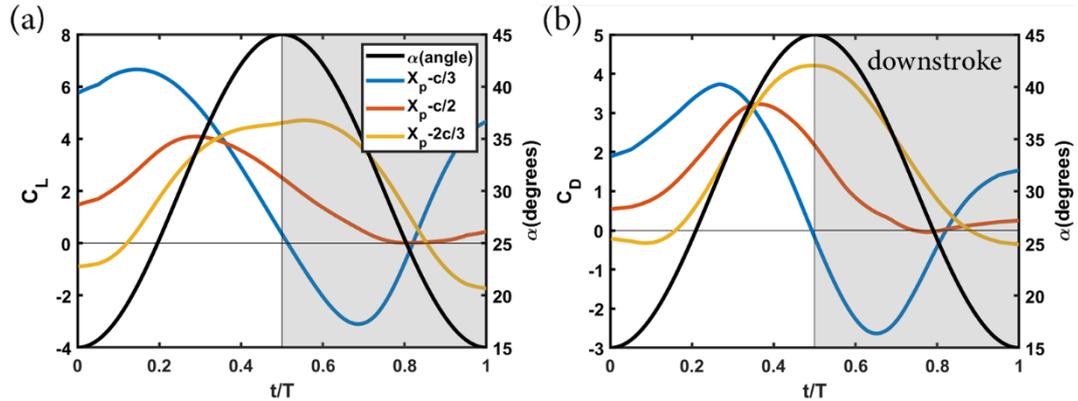

**Figure 22**: Aerodynamic force coefficients on the airfoil compared at three different pivot axis locations ($f^* = 0.5$). (a) plots phase averaged $C_L$ and (b) plots phase averaged $C_D$ against fractional time, respectively.

**Fig. 23a** and **Fig. 23b** plots the variation of phase averaged $C_L$ and $C_D$ against angle of attack for the reduced frequency of 0.5. The width of the $C_L$ and $C_D$ envelope (hysteresis curve) is found to reduce as we move the pivot axis aftward. Unlike our observation at the reduced frequency of 0.1, we find that the $C_L$ and $C_D$ oscillate between the positive and negative values. This indicates that the airfoil experiences drag and thrust, along the freestream direction, and lift and downforce, normal to the freestream direction, at different phases in the pitching motion.

Similar to our observation at $f^* = 0.1$, we have a phase lag in the aerodynamic forces as we move the pivot axis aftward. However, the phase lag at the reduced frequency of 0.5 is more significant. Same observation can be made with respect to the peak forces on the airfoil. At low frequencies ($f^* = 0.1$), the peak values of the aerodynamic forces vary under 5% as the pivot axis changes, whereas at the reduced frequency of 0.5, the change lie between 15.10% and 62.77% . **Section. IIID** details more trends observed in the aerodynamic forces.

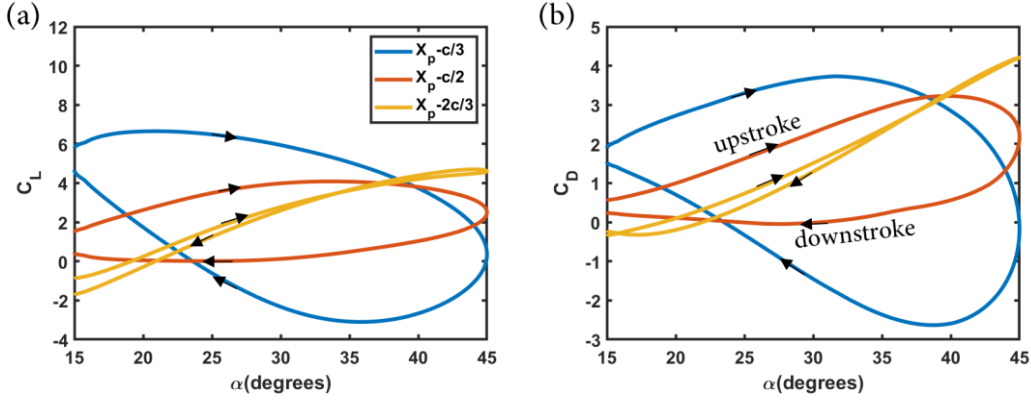

**Figure 23**: Aerodynamic force coefficients on the airfoil compared at three different pivot axis locations ($f^* = 0.5$). (a) plots phase averaged $C_L$ and (b) plots phase averaged $C_D$ against incidence angle, respectively.

The source of the observed phase lag can again be attributed to the change in the effective incidence angle with shift in the pivot axis location. As we move the pivot axis aftward, the effective incidence angle lags, which results in similar trends observed in aerodynamic forces and LEV evolution. Effective angle of attack for a given pivot axis location is a function of pitching frequency. At the reduced frequency of 0.5, the effective incidence for one-third chord pitching, tends to lead mid chord pitching by 0.097T, while two-third chord pitching lags behind mid chord pitching by 0.097T. (More Details in **Appendix. A1**)

## C. Effect of Changing Pitching frequency

In this section, we will explore the effect of pitching frequency on the aerodynamic forces and their sensitivities at different pivot axis locations.

### 1. One-third chord pitching

**Fig. 24** shows the variation of the phase averaged aerodynamic force coefficients along the pitching cycle for the pivot axis location of c/3. **Fig. 24a** and **Fig. 24b** plots $C_L$ and $C_D$ curves against fractional time, respectively. Both the curves have higher amplitude at the reduced frequency of 0.5. There is a 45.44% increment in the maximum drag coefficient and a 92.24% increment in the maximum lift coefficient as the reduced frequency changes from 0.1 to 0.5. Curves corresponding to the reduced frequency of 0.5, oscillate between positive and negative values (Lift and Drag is both positive and negative), whereas the curves corresponding to reduced frequency of 0.1, are restricted to positive values. $C_L$ and $C_D$ curves for both the frequencies tend to lead the motion signal (AOA). Curves corresponding to the reduced frequency of 0.5 reach its maxima first followed by the curves corresponding to the reduced frequency of 0.1. Irrespective of the frequency, $C_{Lmax}$ is found to be greater than $C_{Dmax}$ and this difference is more evident at high frequencies ($f^* = 0.5$).

**Fig. 24c** and **Fig. 24d** plots $C_L$ and $C_D$ curves against angle of attack, respectively. The curves ($C_L$ and $C_D$) corresponding to the reduced frequency of 0.5 tend to have a much wider envelope (hysteresis loop) than those corresponding to reduced frequency of 0.1.

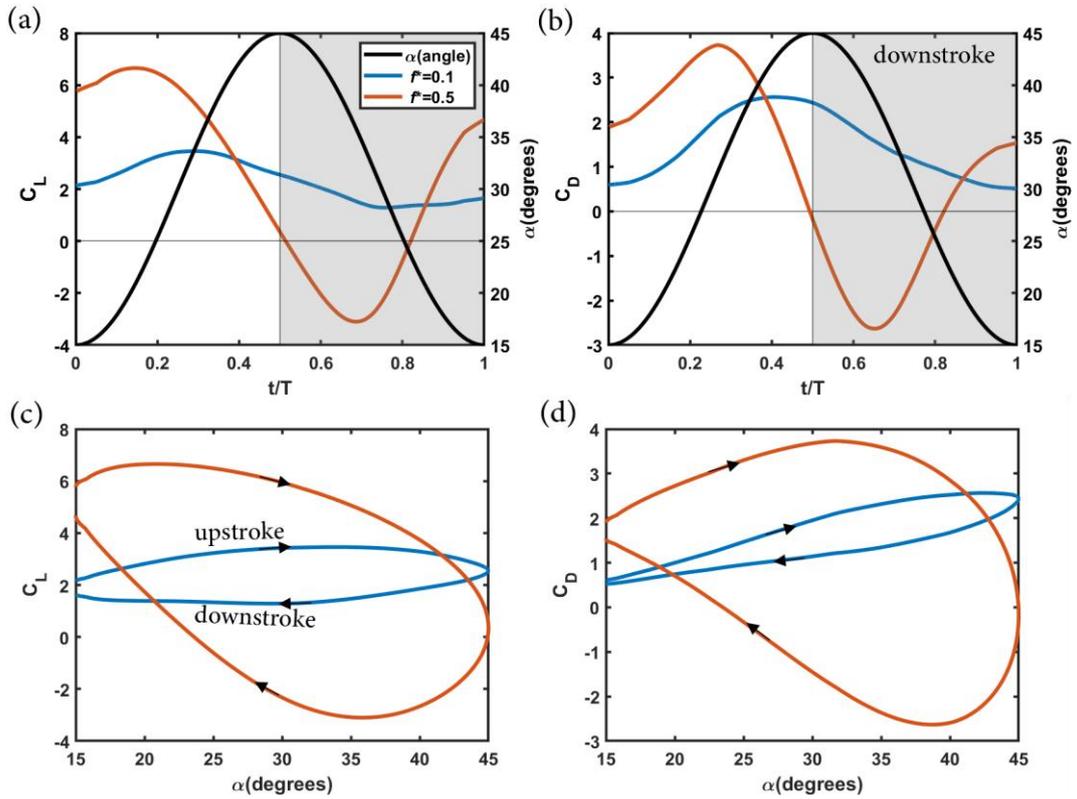

**Figure 24**: Aerodynamic force coefficients on the airfoil compared at two different reduced frequencies (0.1 and 0.5) for the pivot axis location of c/3. (a) and (b) plots phase averaged $C_L$ and $C_D$ respectively against fractional time. (c) and (d) plots phase averaged $C_L$ and $C_D$ respectively against incidence angle.

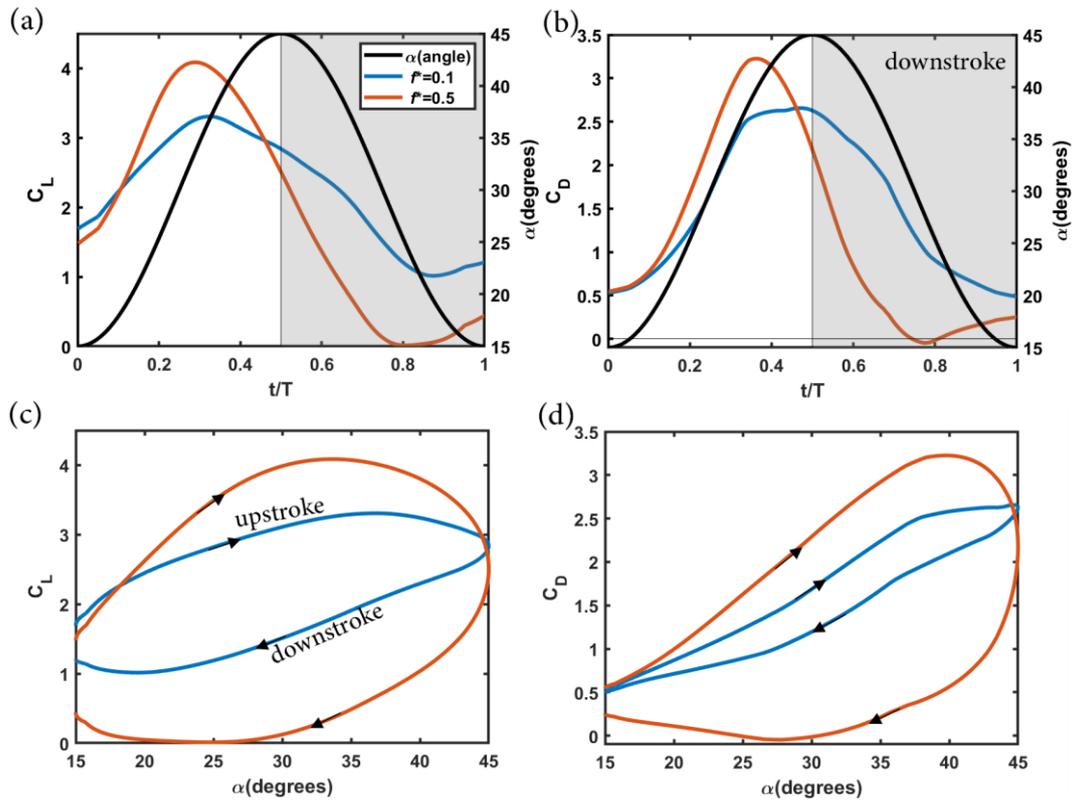

**Figure 25**: Aerodynamic force coefficients on the airfoil compared at two different reduced frequencies (0.1 and 0.5) for the pivot axis location of c/2. (a) and (b) plots phase averaged $C_L$ and $C_D$ respectively against fractional time. (c) and (d) plots phase averaged $C_L$ and $C_D$ respectively against incidence angle.

## 2. Mid chord pitching

**Fig. 25** shows the variation of the phase averaged aerodynamic force coefficients along the pitching cycle for mid chord pitching. **Fig. 25a** and **Fig. 25b** plots $C_L$ and $C_D$ curves against fractional time, respectively. Similar to the observation made for one-third chord pitching, both the curves have higher amplitude at the reduced frequency of 0.5. There is a 21.42% increment in the maximum drag coefficient and a 23.54% increment in the maximum lift coefficient as the reduced frequency changes from 0.1 to 0.5. $C_L$ and $C_D$ curves for both the frequencies tend to lead the motion signal (AOA) and curves corresponding to reduced frequency of 0.5, tend to reach its maxima first followed by the curves corresponding to reduced frequency of 0.1. However, unlike the case with the pivot axis at c/3, the $C_L$ and $C_D$ curves for both the frequencies are restricted to positive values (Phase averaged $C_D$ at $f^* = 0.5$ is negative for a very short span of fractional time during the downstroke motion).

For both the frequencies, $C_{Lmax}$ is found to be greater than $C_{Dmax}$, but the difference is not as significant as that observed when the airfoil is pivoted at c/3 and increasing the pitching frequency does not seem to have a large influence on this difference.

**Fig. 25c** and **Fig. 25d** plots $C_L$ and $C_D$ curves against angle of attack, respectively. The curves ($C_L$ and $C_D$) corresponding to the reduced frequency of 0.5 tend to have a much wider envelope than the curves corresponding to $f^* = 0.1$. The curves corresponding to the reduced frequency of 0.5, almost completely enclose those corresponding to the reduced frequency of 0.1 inside them.

## 3. Two-third chord pitching

**Fig. 26** plots the variation of the phase averaged aerodynamic force coefficients along the pitching cycle for the pivot axis location of 2c/3. From **Fig. 26a** and **Fig. 26b**, the $C_L$ curve corresponding to the reduced frequency of 0.5, tend to lag behind the motion signal whereas the $C_D$ curve for the same frequency is almost in-phase with the angle of attack. The $C_L$ and $C_D$ curves corresponding to the reduced frequency of 0.1 leads the motion signal. The amplitude of the curves ($C_L$ and $C_D$) corresponding to the reduced frequency of 0.5, are much larger than that at 0.1. There is a 53.87% increment in the maximum drag coefficient and a 41.71% increment in the maximum lift coefficient as the reduced frequency changes from 0.1 to 0.5. The $C_L$ and $C_D$ values oscillate between positive and negative values at the reduced frequency of 0.5, whereas they are restricted to positive values at the reduced frequency of 0.1.

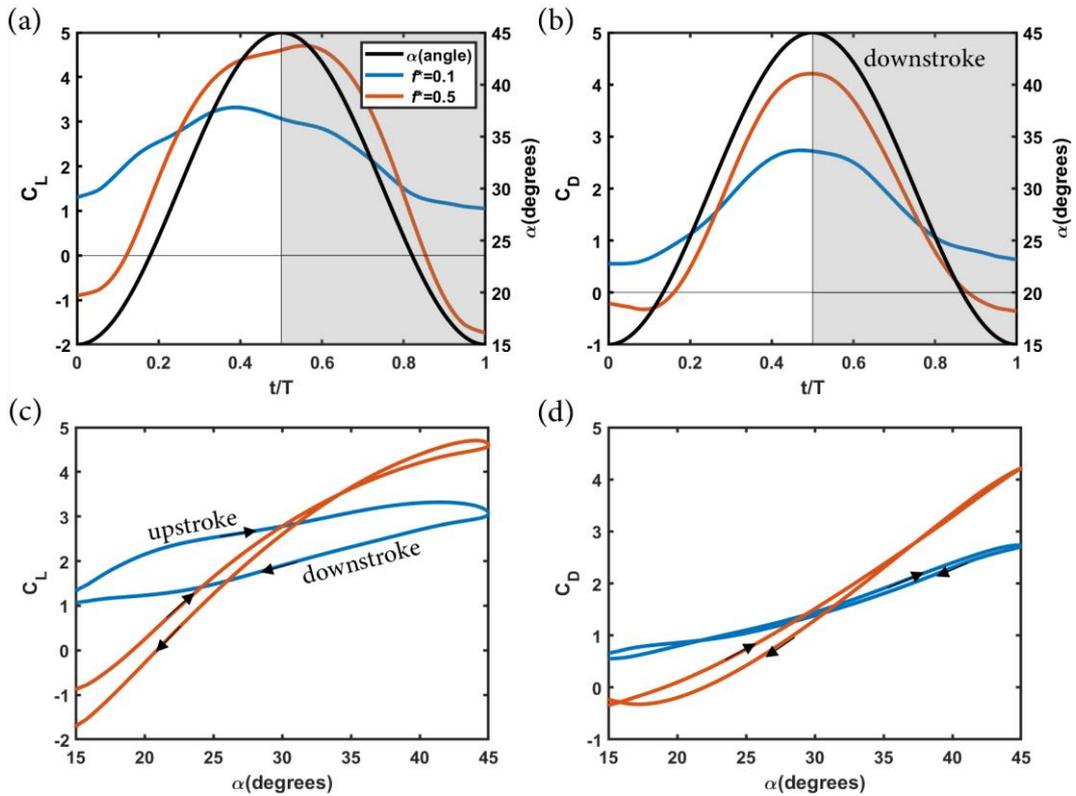

**Figure 26**: Aerodynamic force coefficients on the airfoil compared at two different reduced frequencies (0.1 and 0.5) for the pivot axis location of 2c/3. (a) and (b) plots phase averaged $C_L$ and $C_D$ respectively against fractional time. (c) and (d) plots phase averaged $C_L$ and $C_D$ respectively against incidence angle.

Similar to the observation made for mid-chord pitching, for both frequencies, $C_{Lmax}$ is greater than $C_{Dmax}$ with only a small difference between them, compared to that observed for one-third chord pitching. From **Fig. 26c** and **Fig. 26d**, the $C_L$ envelope at the reduced frequency of 0.5 is found to be wider than that observed at $f^* = 0.1$, but the $C_D$ envelope widths of both the frequencies are comparable, even though their peaks differ significantly.

### D. Aerodynamic forces on the airfoil

We have explored the flow evolution at the reduced frequencies of 0.1 and 0.5 in **Sections III.A** and **III.B**, respectively. At the reduced frequency of 0.1, the evolution of dominant vortices is found to happen in two possible patterns during the downstroke motion. Even within an evolution type, the vortical strengths can vary from cycle to cycle. Thus, the peak and the average aerodynamic forces vary in each cycle. At high frequencies ($f^* = 0.5$), multiple LEVs exist on the wing forming a cluster. Number of vortical structures on the wing vary from cycle to cycle and this variation is expected in the aerodynamic forces as well.

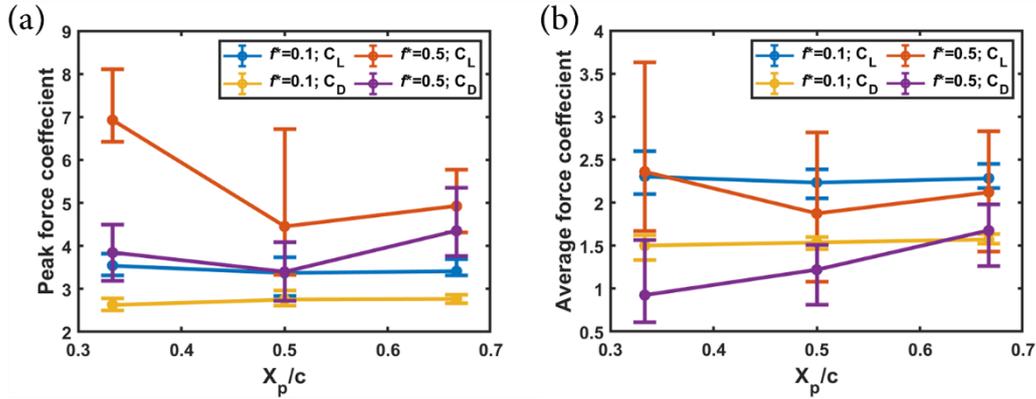

**Figure 27**: Variation in the averaged peak (a) and mean (b) values of the force coefficient ($C_L$ and $C_D$) with change in pivot axis location at the reduced frequencies of 0.1 and 0.5. Maximum deviations observed in the coefficients for 10 pitching cycles is indicated with whiskers.

**Fig. 27a** and **Fig. 27b** plots the variation in the averaged peak and mean values of the force coefficients ($C_L$ and $C_D$) in 10 pitching cycles, respectively, against pivot axis location at the reduced frequencies of 0.1 and 0.5. Note that the average of the peak values of 10 pitching cycles is not the same as the maxima of the phase averaged plot, but the average of the mean of 10 pitching cycles is equal to the mean of the phase averaged variation. The minimum and maximum variations observed in these coefficients in 10 pitching cycles are indicated using whiskers. On an average, variations in the peak values are found to be greater than the corresponding variations in the mean coefficients. These deviations are also found to be greater at high frequencies ($f^* = 0.5$). This can be attributed to the significant difference in the evolution of vortices in each cycle at the reduced frequency of 0.5 (**Fig. 18**). The averaged peak of the force coefficients increases with increase in pitching

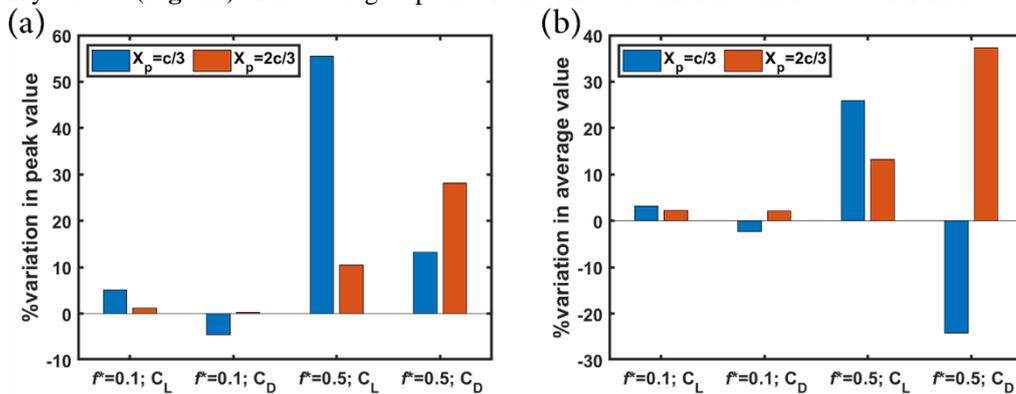

**Figure 28**: Percentage variation in the averaged peak (a) and mean (b) values of the force coefficient ($C_L$ and $C_D$) in one-third and two-third chord pitching with respect to mid-chord pitching

frequency, for all pivot axis locations. This can be attributed to rise in the inertial component of force and change in the vorticity evolution. (Aerodynamic forces on a pitching airfoil can be broken down into two components, inertial component which is a strong function of pitching frequency and the circulatory component which depends on the vorticity evolution in the flow field).

An interesting observation is that, even though the averaged peak values rise with pitching frequency, averaged mean values do not follow the trend for all pivot axis location. Only for one-third chord pitching, we observe the rise in the averaged mean lift coefficient with increase in pitching frequency. For all other pivot axis location, increase in frequency, tends to reduce this value. Similarly, the rise in averaged mean drag coefficient with pitching frequency is observed only at the pivot axis location of 2c/3.

Even though the trends in the averaged peak and mean values of the force coefficients are evident from **Fig. 27,** such generalized trends are not visible with change in pivot axis location. **Fig. 28** plots the percentage change in the averaged peak and mean values of the force coefficients for one-third and two-third chord pitching with respect to mid-chord pitching (baseline case), at the reduced frequencies of 0.1 and 0.5. The plot presents the percentage change observed in the force coefficients, as we move the pivot axis by c/6 ahead and behind the mid-chord axis. Except for the averaged mean value of the drag coefficient in one-third chord pitching, mid-chord pitching is found to have the least averaged peak and mean in $C_L$ and $C_D$ at the reduced frequency of 0.5. The same trend is observed in the peak and mean values of the averaged lift coefficient at $f^* = 0.1$. However, this is not the case with the drag coefficient. Averaged peak and mean of the $C_D$ is found to increase with distance of the pivot axis from the leading edge, at the reduced frequency of 0.1.

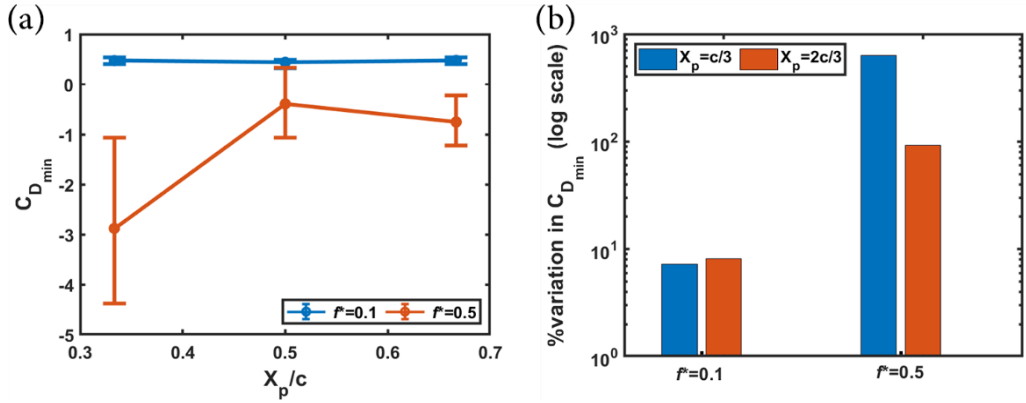

**Figure 29**: (a) plots the variation of the averaged value of $(C_D)_{min}$ at the reduced frequencies of 0.1 and 0.5. The whiskers show the deviation on either side of the averaged minimum observed in 10 pitching cycles. (b) plots the percentage variation in the averaged $(C_D)_{min}$ for one-third and two-third chord pitching, with respect to that observed in mid-chord pitching.

**Fig. 29a** plots the averaged minimum drag coefficient at the reduced frequencies of 0.1 and 0.5, against pivot axis location. The whiskers show the variation observed in $(C_D)_{min}$ in 10 pitching cycles with respect to the averaged minimum drag coefficient. (Once again, average of the minimum drag coefficient of 10 pitching cycles is not the same as the minimum of the phase averaged drag coefficient) The estimates are found to be negative at the reduced frequency of 0.5, indicating the existence of thrust at certain phases in the pitching cycle. However, this is not observed at the reduced frequency of 0.1. **Fig. 29b** plots the percentage change in the averaged $(C_D)_{min}$ as we move the pivot axis location ahead and behind the mid-chord location by c/6. At $f^* = 0.5$, averaged $(C_D)_{min}$ is found to be least for one-third chord pitching and highest for mid-chord pitching, whereas at the reduced frequency of 0.1, averaged $(C_D)_{min}$ is found to be least for mid-chord pitching and highest for two-third chord pitching. There is only a 10% variation in the averaged minimum drag values as we change the pivot axis location at the reduced frequency of 0.1, whereas the change is drastic at $f^* = 0.5$. Moving the pivot axis ahead of the mid-chord by c/6 increases the average $(C_D)_{min}$ by 634% and moving the pivot axis behind mid-chord by c/6 enhances it by 92%, both in the negative sense.

## IV. Conclusion

We have numerically studied the dynamics of high amplitude continuous pitching airfoils using a finite volume based sharp interface immersed boundary solver. The evolution of leading edge vortices has been investigated in detail at the reduced frequencies of 0.1 and 0.5. Vortex tracking (LEV-TEV tracking) was used along the pitching cycle to obtain a quantitative estimate of flow evolution. These estimates were further used to explain the flow features and its implications on the aerodynamic forces. This work mainly focused on studying the effects of reduced frequency and pivot axis location on the vortex evolution and aerodynamic forces.

The leading edge vortex dynamics of high amplitude continuously pitching airfoils are highly non-linear, non-cyclical. Observations over 10 pitching cycles demonstrated that some pattern can be observed in terms of leading edge vortex disintegration into secondary vortices at low frequencies or the vortex clustering phenomena at the high frequencies which were quantitatively analyzed in the present study.

In terms of aerodynamic force characterictics, at high frequencies ($f^* = 0.5$), they are found to be extremely sensitive to pivot axis location. Both peak amplitude and phase of the aerodynamic forces, changed significantly with pivot axis. LEV evolution and the aerodynamic force characteristics were found to be delayed to higher fractional times as the pivot axis is moved aftward. The observed phase shifts were justified based on the effective incidence angle. This phase lag is observable even at low frequencies. However, the peak value of the forces is found, not to alter significantly with change in pivot axis location at low frequencies ($f^* = 0.5$). Changes in the peak and mean of the phase averaged aerodynamic forces were compared at reduced frequencies of 0.1 and 0.5 for three different pivot axis location (c/3, c/2 and 2c/3), to explore their sensitivities. Increasing the reduced frequency was found to increase the peak amplitude, irrespective of the pivot axis location. However, phase shift did not show a monotonic change with pitching frequency.

## IV. Appendix

### A1: Effective Angle of attack at different pivot axis locations

Zhenyao[28] and Tian[29] showed that a kinematic parameter, the effective angle of attack, can be used to trace out the source of the phase lag observed in the aerodynamic forces with change in pivot axis location. Tian's Studies[29] on pitching airfoils show that pitching about any point on the airfoil can be broken down into pitching about a baseline point at the same frequency, and a plunging motion, which in itself is a function of the baseline pitching. Thus, we can resolve the pitching about the one-third and two-third chord axis as a function of mid chord pitching. The method helps in expressing the effective incidence angle for pitching about any pivot axis with respect to that experienced in mid chord pitching.

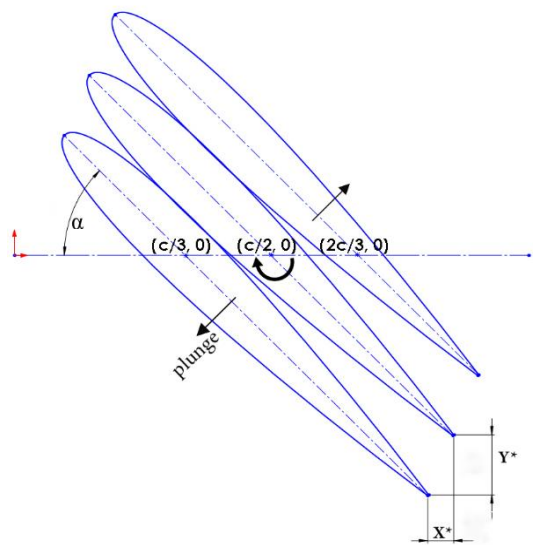

**Figure 30**: Pitching about one-third and two-third axis expressed in terms of mid-chord pitching.

**Fig. 30** shows a schematic of pitching about the one-third and two-chord axis expressed in terms of mid-chord pitching. From the figure, we can see that, at an instance during the motion (when the angle of attack imposed by pitching is $\alpha$), the orientation of the foil in one-third chord pitching, can be expressed based on the incidence angle at mid-chord pitching defined at that time instance, and displacements along ($X^*$) and normal ($Y^*$) to the freestream direction ($X^*$ and $Y^*$ are zero at $\alpha = 0^o$). Thus, the pitching motion about the one-third chord axis can be expressed as pitching about the mid chord axis along with a plunging motion, with velocity components along and normal to the freestream direction.

Similarly, two-third chord pitching or pitching about any point ($x$) on the airfoil, can also be broken down into mid-chord pitching and a plunging motion. The formulation for displacements: $X^*$ and $Y^*$, for a pitching about an axis which is at a distance of $x$ from the leading edge is given by,

$$X^* = \left(\frac{c}{2} - x\right)(cos\alpha - 1) \qquad Y^* = -\left(\frac{c}{2} - x\right)sin\alpha \qquad (9)$$

Velocities along ($U$) and normal ($V$) to the freestream directions due to the added plunging motion can be obtained by differentiating the above expressions for displacements with respect to time. Note that these velocity components are function of the instantaneous pitch angle ($\alpha$). The added velocities, along with the freestream velocity dictates the effective angle of attack ($\alpha_{eff}$).

$$\alpha_{eff} = \alpha + tan^{-1}\left[\frac{V}{U + U_\infty}\right] \qquad (10)$$

Effective incidence angle plotted for one-third and two-third chord pitching is shown in **Fig. 31.** We observe that, with respect to mid-chord pitching, the effective incidence angle leads by 0.023T and 0.097T when the airfoil is pivoted at c/3 (c/6 ahead of the mid-chord location), at the reduced frequencies of 0.1 and 0.5 respectively. A proportionate lag is observed when the airfoil is pivoted at a distance of c/6 behind the mid-chord ($X_P = 2c/3$). The phase shift in the effective incidence angle explains the corresponding shifts in the aerodynamic forces and accounts for the delay in the LEV evolution.

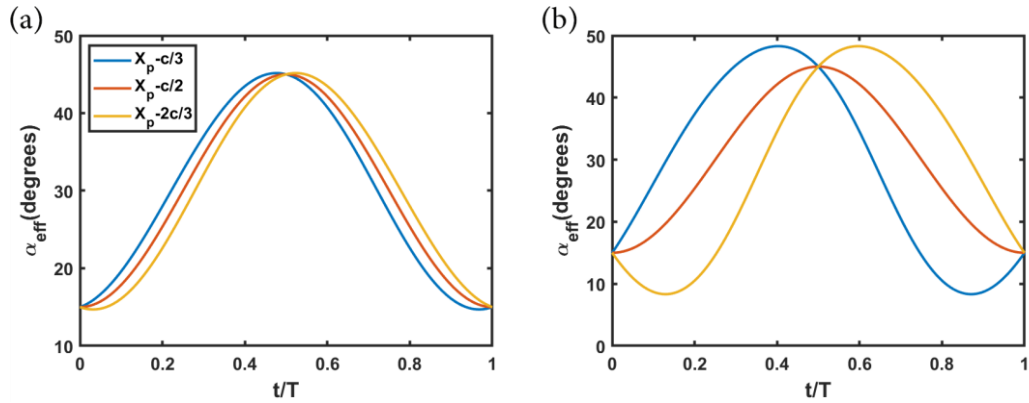

**Figure 31**: Effective incidence angle for one-third and two-third chord pitching with respect to mid-chord pitching, at (a) $f^* = 0.1$ and (b) $f^* = 0.5$